\documentclass{cpbtex}
\usepackage[utf8]{inputenc}

\title{Restricted Boltzmann Machine, recent advances and mean-field theory\thanks{AD was supported by the Comunidad de Madrid and the Complutense University of Madrid (Spain) through the Atracción de Talento program (Ref. 2019-T1/TIC-13298).}}
\author{Aurélien Decelle$^{1,2}$\thanks{Corresponding author. E-mail:~adecelle@ucm.es} \ and \ Cyril Furtlehner$^{2}$ \\
$^{1}${Departamento de F\'isica T\'eorica I, Universidad Complutense, 28040 Madrid, Spain}\\  % The line break was forced via \\
$^{2}${TAU team INRIA Saclay \& LISN Université Paris Saclay, Orsay 91405, France}} % The line break was forced via \\

\date{}
\usepackage{graphicx}
\usepackage{bm}
\usepackage{amsmath}
\usepackage{amssymb}
\usepackage{amsfonts}
\usepackage{caption}
\usepackage{subcaption}
\usepackage{soul}
\usepackage{ulem}
\usepackage{xcolor}

\begin{document}
\maketitle
% \tableofcontents
\begin{abstract}
    This review deals with Restricted Boltzmann Machine (RBM) under the light of statistical physics. The RBM is a classical family of Machine learning (ML) models which played a central role in the development of deep learning. Viewing it as a Spin Glass model and exhibiting various links with other models of statistical physics, we gather recent results dealing with mean-field theory in this context. First the functioning of the RBM can be analyzed via the phase diagrams obtained for various statistical ensembles of RBM leading in particular to identify a {\it compositional phase} where a small number of features or modes are combined to form complex patterns. Then we discuss recent works either able to devise mean-field based learning algorithms; either able to reproduce generic aspects of the learning process from some {\it ensemble dynamics equations} or/and from linear stability arguments.
\end{abstract}


% \href{http://cpb.iphy.ac.cn/EN/column/item208.shtml}{PACS codes} should be provided

\section{Introduction}

During the last decade, machine learning has experienced a rapid development, both in everyday life with the incredible success of image recognition used in various applications, and in research\ucite{goodfellow2016deep,mehta2019high} where many different communities are now involved. This common effort involves fundamental aspects such as why it works or how to build new architectures and at the same time a search for new applications of machine learning to other fields, like for instance improving biomedical images segmentation\ucite{ronneberger2015u} or detecting automatically phase transitions in physical system\ucite{carrasquilla2017machine}.  Machine learning classical tasks are divided into at least two big categories: supervised and unsupervised learning (putting aside reinforcement learning and the more recently introduced approach of self-supervised learning). Supervised learning consists in learning a specific task --- for instance recognizing an object on an image or a word in a speech--- by giving the machine a set of samples together with the correct answer and correcting the prediction of the machine by minimizing a well-design and easy computable loss function. Unsupervised learning consists in learning a representation of the data given an explicit or implicit probability distribution, hence adjusting a likelihood function on the data. In this latter case, no label is assigned to the data and the result depends thus solely on the structure of the considered model and of the dataset.

In this review, we are interested in a particular model: the Restricted Boltzmann Machine (RBM). Originally called Harmonium~\ucite{Smolensky} or product of experts~\ucite{hinton2002training}, RBMs were designed\ucite{ackley1985learning} to perform unsupervised tasks even though they can also be used to accomplish supervised learning in some sense. RBMs are part of what is called generative models which aim to learn a latent representation of the data in order to later be used to generate statistically similar new data ---but different from those of the training set. There are Markov Random Fields (or Ising model for physicists), that were designed as a way to automatically interpret an image using a parallel architecture including a direct encoding of the probability of each ``hypothesis'' (latent description of a small portion of an image). Later on, RBMs started to take an important role in the Machine Learning community, when a simple learning algorithm introduced by Hinton et al.\ucite{hinton2002training}, the contrastive divergence (CD), managed to learn a non trivial dataset such as MNIST\ucite{lecun1998gradient}. It was in the same period that RBMs became very popular in the ML community for its capability to pre-train deep neural networks (for instance deep auto-encoder), in a layer wise style. And, it was then showed that RBMs are universal approximator\ucite{le2008representational} of discrete distributions, that is, an arbitrary large RBM can approximate arbitrarily well any discrete distribution (which led to many rigorous results about the modelization mechanism of RBMs\ucite{montufar2016restricted}). In addition, RBMs offer the possibility to be stacked to form a multi-layer generative model known as a deep Boltzmann machine (DBM)\ucite{salakhutdinov2009deep}. In the more recent years, RBMs continued to attract scientific interest. Firstly because it can be used on continuous or discrete variable very easily\ucite{krizhevsky2009learning,MuTa,cho2011improved,yamashita2014bernoulli}. Secondly, because the possible interpretations of the hidden nodes can be very useful\ucite{hjelm2014restricted,hu2018latent}. Interestingly, in some cases, more elaborate methods such as GAN\ucite{goodfellow2014generative} are not working better\ucite{yelmen2019creating}. Finally it can be used for other tasks as well such as classification or representation learning\ucite{ZHANG20181186}. Besides all these positive aspects, the learning process itself of the RBM remains  poorly understood. The reasons are twofold: firstly, the gradient can be computed only in an approximated way as we will see; secondly, simple changes may have terrible impact on the learning or, messed up completely with the other meta-parameters. For instance making a naive change of variable in the MNIST dataset\ucite{cho2011enhanced,tang2011data} can affect importantly the training performance\footnote{In MNIST, it is usual to consider binary variable $\{0,1\}$ to describe the dataset. Taking instead $\{\pm1\}$ naively will affect dramatically the learning of the RBM.}. Another example, when varying the number of hidden nodes, keeping the other mete-parameters fixed, will affect not only the representational power of the rbm but also the learning dynamics itself.

The statistical physics community, on its side, has a long tradition of studying inference and learning process with its own tools. Using idealized inference problems, it has managed in the past to shed light on the learning process of many ML models. For instance, in the Hopfield model\ucite{hopfield1982neural,AmGuSo1,AmGuSo2,AmGuSo3}, a retrieval phase was characterized where the maximum number of patterns that can be retrieved can be expressed as a function of the temperature. Another example is the computation of the storage capacity of the Perceptron\ucite{rosenblatt1958perceptron} on synthetic datasets\ucite{gardner1988space,Derrida-Gardner}. In these approaches, the formalism of statistical physics explains the macroscopic behavior of the model in term of its position on a phase diagram in the large size limit.

From a purely technical point of view, the RBM can be seen for a physicist as a disordered Ising model on a bipartite graph. Yet, the difference with respect to the usual models that are studied in statistical physics is that the phase diagram of a trained RBM involves a highly non-trivial coupling matrix where the components are correlated as a result of the learning process. These dependencies make it non-trivial to adapt classical tools from statistical mechanics, such as the replica theory\ucite{mezard1987spin}. We will illustrate in this article how methods from statistical physics still have helped to characterize both the equilibrium phase of an idealized RBM where the coupling matrix has a structured spectrum, and how the learning dynamics can be analyzed in some specific regimes, both results being obtained with traditional mean-field approaches.

The paper is organized as follows. We will first give the definition of the RBM and review the typical learning algorithm used to train the model in Section~(\ref{sec:def}). Then, in Section~(\ref{sec:link_rbm}), we will review different types of RBMs by changing the prior on its variables and show explicit links with other models. In Section~(\ref{sec:phase_diag}), we will review two approaches that characterize the phase diagram of the RBM and in particular its {\it compositional phase}, based on two different hypothesis over the structure of the parameters of the model. Finally, in Section~(\ref{sec:learning_rbm}) we will show some theoretical development helping to understand the formation of patterns inside the machine and how we can use the mean-field or TAP equations to learn the model.

\section{Definition of the model and learning equations}
\label{sec:def}
\subsection{Definition of the RBM}
\begin{figure}[ht!]
    \centerline{\resizebox*{0.7\textwidth}{!}{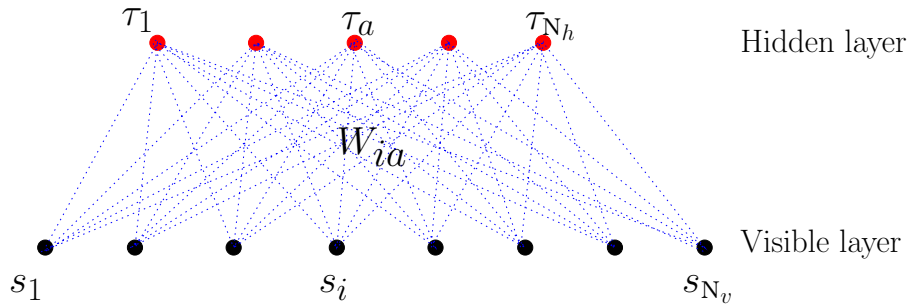}}
    \caption{\label{fig:rbm} bipartite structure of the RBM.}
\end{figure}

The RBM is an Ising model (or equivalently, a Markov random field), defined on a bipartite graph structure over two layers of variables: the visible nodes $s_i$, for $i=1,\dots,N_v$ and the hidden nodes $\tau_a=1,\dots,N_h$, denoting $N_v$ and $N_h$ the number of visible and hidden nodes respectively. In the following, we will use $i,j,k,\dots$ to enumerate the visible variables and $a,b,c,\dots$ for the hidden ones. No connection between any pair of visible or hidden nodes occurs . Hence, we will call $\bm{w}$ the coupling or weight matrix and denote its elements as $w_{ia}$ since no other interactions are present (such as $w_{ij}$ or $w_{ab}$). In addition to the pairwise coupling matrix $\bm{w}$, each visible and hidden node can have a local magnetic field, or local bias (we will refer to it as bias in the rest of the article), respectively named $\theta_i$ and $\eta_a$. We can introduce the following Hamiltonian
\begin{equation}\label{eq:H}
    \mathcal{H}[\bm{s},\bm{\tau}] = -\sum_{ia} s_i w_{ia} \tau_a - \sum_i \theta_i s_i - \sum_a \eta_a \tau_a,
\end{equation}
from which we define a Boltzmann distribution
\begin{equation*}
    p(\bm{s},\bm{\tau}) = \frac{1}{Z}\exp(-\mathcal{H}[\bm{s},\bm{\tau}]).
\end{equation*}
where $Z$ is given by
\begin{equation*}
  Z = \sum_{\{\bm{s}\},\{\bm{\tau}\}} \exp(-\mathcal{H}[\bm{s},\bm{\tau}]).
\end{equation*}
\noindent The structure of the RBM is represented on Figure~\ref{fig:rbm}

\noindent where the visible nodes are represented by black dots, the hidden nodes by red dots and the weight matrix by blue dotted lines.

\noindent The benefit of having a bipartite structure is that, when keeping fixed an entire layer, in our case all the visible or all the hidden nodes, the variables of the other layer become statistically independent. In other words, the measure $p(\bm{s}|\bm{\tau})$ and $p(\bm{\tau}|\bm{s})$ factorizes over the visible/hidden nodes respectively. This is an important property to keep in mind since it will be used in the learning procedure of the model. We will see that this property is widely used during the learning in order to draw new samples using a Monte-Carlo Markov Chain (MCMC) by Gibbs sampling. 

Historically, the RBM was first defined with binary $\{0,1\}$ variables for both the visible and the hidden nodes in line with the sigmoid activation function of the perceptron, hence being directly intepretable as spin-glass model of statistical mechanics. A more general definition is considered here by introducing a prior distribution function for both the visible and hidden variables, allowing us to consider discrete or continuous variables. This generalization will allow us to see the links between RBMs and other well-known models of machine learning. From now on we will write all the equations for the generic case using the notation $q_v(\sigma)$ and $q_h(\tau)$ to indicate an arbitrary choice of ``prior'' distribution. Averaging over the RBM measure corresponding to Hamiltonian~(\ref{eq:H}) will then be denoted by
\begin{equation}
    \langle f(\bm{s},\bm{\tau}) \rangle_\mathcal{H} = \sum_{\{s,\tau\}} p(\bm{s},\bm{\tau})  f(\bm{s},\bm{\tau})
\end{equation}
where here $\Sigma$ can represent both discrete sums or integrals and with the RBM distribution defined from now on as
\begin{equation}
p(\bm{s},\bm{\tau}) = \frac{1}{Z} q_v(\bm{s}) q_h(\bm{\tau}) \exp(-\mathcal{H}[\bm{s},\bm{\tau}]).
\label{eq:bolt}
\end{equation}

\noindent It is worth mentioning that, the choice of the prior distribution can be rephrased in terms of an activation function on the conditioned distribution over the visible or hidden variables. Therefore, when specifying a prior distribution, we will systematically indicate the corresponding activation function for the hidden layer, that is $p(\bm{\tau}|\bm{s})$, which is obtained using the Bayes theorem
\begin{equation*}
  p(\bm{\tau}|\bm{s}) = \frac{p(\bm{s},\bm{\tau})}{\sum_{\tau} p(\bm{s},\bm{\tau})} = \frac{q_h(\bm{\tau})  \exp(-\mathcal{H}[\bm{s},\bm{\tau}])}{\sum_{\{\bm{\tau}\}}q_h(\bm{\tau})  \exp(-\mathcal{H}[\bm{s},\bm{\tau}])}
\end{equation*}
Before entering more into the technical details about the RBM, it is important to recall that it has been designed as a ``learnable'' generative model in practice. In that sense, the usual procedure is to feed the RBM with a dataset, tune its parameter $w$, $\theta$ and $\eta$ such that the equilibrium properties of the learned RBM reproduce faithfully the correlations (or the patterns) present in the dataset. In other words, it is expected that the learned model is able to produce new data statistically similar but distinct from the training set. To do so, the classical procedure is to proceed with a stochastic gradient ascent (to be explained in Section~\ref{sec:stochgrad}) of the likelihood function that can be easily expressed. Usually the learning of ML models involves the minimization of a loss function which happens here to be minus the log likelihood, thus in the following we will refer to Stochastic Gradient Descent (SGD) instead. First, consider a set of datapoints $\{s_i^{(d)}\}$, where $d=1,\dots,M$ is the index of the data. The log-likelihood is given by
\begin{align}
    \mathcal{L} & = \frac{1}{M} \sum_{d=1}^M \log\left( \sum_{\{\bm{\tau}\}} p(\bm{s}^{(d)},\bm{\tau})\right)  = \frac{1}{M} \sum_{d=1}^M \log\left( p(\bm{s}^{(d)} )\right) \nonumber \\
    & =\frac{1}{M}  \sum_{d=1}^M\left[ \log\left(\sum_{\tau} q_v(\bm{s}^{(d)})q_h(\bm{\tau})\exp\bigl(-\mathcal{H}[\bm{s}^{(d)},\bm{\tau}]\bigr)\right)\right] - \log(Z)  \nonumber \\
    & =\frac{1}{M}  \sum_{d=1}^M\left[ \sum_i \theta_i s_i^{(d)} + \log\left( q_v(\bm{s}^{(d)}) \right) + \sum_a \log\left(\sum_{\tau_a} q_h(\tau_a) \exp\bigl(\sum_{i} s_i^{(d)} w_{ia} \tau_a + \eta_a \tau_a\bigr)\right) \right] - \log(Z) \nonumber
\end{align}

\noindent The gradient w.r.t. the different parameters will then take a simple form. Let us detail the computation of the gradient w.r.t. the weight matrix. By deriving the log-likelihood w.r.t. the weight matrix we get
\begin{align}
    \frac{\partial \mathcal{L}}{\partial w_{ia}} &= \frac{1}{M} \sum_{d=1}^M \frac{\sum_{\tau_a} q_h(\tau_a) s_i^{(d)} \tau_a e^{\sum_{i} s_i^{(d)} w_{ia} \tau_a + \eta_a \tau_a}}{ \sum_{\tau_a} q_h(\tau_a) e^{\sum_{i} s_i^{(d)} w_{ia} \tau_a + \eta_a \tau_a}}  - \langle s_i \tau_a \rangle_\mathcal{H} \nonumber \\
    & = \frac{1}{M} \sum_{d=1}^M  s_i^{(d)} \sum_{\tau_a} \tau_a p(\tau_a|\bm{s}^{(d)}) - \langle s_i \tau_a \rangle_\mathcal{H} \nonumber \\
    & = \langle s_i \tau_a \rangle_{\rm data} -  \langle s_i \tau_a \rangle_\mathcal{H} \label{eq:sgd:w}
\end{align}

\noindent where we used the following notation
\begin{equation}
    \langle f(\bm{s},\bm{\tau}) \rangle_{\rm data}  = \frac{1}{M} \sum_{d=1}^M \sum_{\{\bm{\tau}\}} f(\bm{s}^{(d)},\bm{\tau}) p(\bm{\tau}|\bm{s}^{(d)}).
\end{equation}

\noindent The gradients for the biases (or magnetic fields) are
\begin{align}
    \frac{\partial \mathcal{L}}{\partial \theta_i} &= \langle s_i\rangle_{\rm data} -  \langle s_i \rangle_\mathcal{H} \label{eq:sgd:theta} \\
    \frac{\partial \mathcal{L}}{\partial \eta_a} &= \langle \tau_a\rangle_{\rm data} -  \langle \eta_a \rangle_\mathcal{H} \label{eq:sgd:eta}
\end{align}

\noindent It is interesting to note that, in expression (\ref{eq:sgd:w}), the gradient is very similar to the one obtained in the traditional inverse Ising problem with the difference that in the inverse Ising  problem the first term (sometimes coined ``positive term'') depends only on the data, while for the RBM, we have a dependence on the model (yet simple to compute). Once the gradient is computed, the parameters of the model are updated in the following way
\begin{align}
    w_{ia}^{(t+1)} &= w_{ia}^{(t)} + \gamma  \frac{\partial \mathcal{L}}{\partial w_{ia}}\Bigr|_{w_{ia}^{(t)},\theta_{i}^{(t)},\eta_a^{(t)}} \\
    \theta_i^{(t+1)} &= \theta_i^{(t)}+ \gamma  \frac{\partial \mathcal{L}}{\partial \theta_{i}}\Bigr|_{w_{ia}^{(t)},\theta_{i}^{(t)},\eta_a^{(t)}} \\ 
    \eta_a^{(t+1)} &= \eta_a^{(t)}+ \gamma  \frac{\partial \mathcal{L}}{\partial \eta_{a}}\Bigr|_{w_{ia}^{(t)},\theta_{i}^{(t)},\eta_a^{(t)}}
\end{align}

\noindent where $\gamma$ called the learning rate tunes the speed at which the parameters are updated in a given direction, the superscript $t$ being the  index of iteration. A continuous
limit of the learning process can be formally defined by considering $t$ real and replacing $t+1$ by $t+dt$, $\gamma$ by $\gamma dt$ and letting $dt\to 0$.

The difficulty to train an RBM resides in the difficulty to compute the second term of  the gradient, the so-called ``negative term'', which represents, in the gradient over the weight matrix, the correlation between a visible node $i$ and a hidden node $a$ under the RBM distribution. Similarly, the gradient over the biases is difficult to compute, where here the negative term is given by the mean value over the visible/hidden nodes. Depending on the value of the parameters of the model (the couplings and the biases), we can either be in a phase where it is easy to sample configurations from  $p(\bm{s},\bm{\tau})$, (usually called paramagnetic phase); either be (if unlucky) in a spin glass phase, where it is exponentially hard to escape from the spurious free energy minima; either be (if lucky) in a "recall" phase where the dominant states correspond to data-like configurations. But even in the latter case, it might be difficult to transit from one state to another one with random jumps if these states are separated by large energy or free energy barriers, as in the Hopfield model for instance.

\subsection{Stochastic Gradient Descent}\label{sec:stochgrad}

Considering the difficulty to use the eq. (\ref{eq:sgd:w}) to learn the model (the computation of the negative term scales exponentially with the system size, and Monte Caro Markov chains (MCMC) can be very slow to converge), an efficient approximative scheme name contrastive divergence\ucite{hinton2002training} (CD) has been developed in order to approximate this term. First of all, the dataset is partitioned into small subsets called minibatches, and the gradient ascent is performed sequentially over all these minibatches in a random order. As a result each gradient step is performed only over a small subset of the whole dataset at a time. In order to estimate the negative term, the principle of CD is to start many Monte-Carlo chains in parallel, as many as the number of samples in a minibatch, and to use each sample of the minibatch as an initial condition for the chain. The idea  being that starting from desired equilibrium configurations and making $k$ steps --- the number of MC steps is coined in the method : CD-k---, we expect to explore nearby configurations representative of the dataset when the machine is learned; if otherwise the chains flow away they will ``teach'' the RBM how to adjust the parameters. The interpretation of CD is that it tends to create a basin of "attraction" centered on the datapoints where nearby configurations will be attractive to these datapoint under the Gibbs dynamics. In practice, starting from a datapoint $\bm{s}^d$ a random configuration of the hidden layer is sampled; in turn  given this a configuration of the visible layer is sampled and so on for $k$ steps. For this we take advantage of the bipartite structure of the model to draw a whole visible or hidden layer at once thanks to the factorization of the conditional distribution $p(\bm{s}|\bm{\tau})$ and $ p(\bm{\tau}|\bm{s})$:
\begin{equation}
    \bm{s}^d \rightarrow \bm{\tau}_0 \sim p(\bm{\tau}|\bm{s}^d) \rightarrow \bm{s}_1 \sim p(\bm{s}|\bm{\tau}_0) \rightarrow \dots  \rightarrow \bm{s}_k \sim p(\bm{s}|\bm{\tau}_{k-1}) \rightarrow \bm{\tau}_k \sim p(\bm{\tau}|\bm{s}_k)
\end{equation}

\noindent finally $\bm{s}_k$ and $\bm{\tau}_{k}$ are used to estimate the negative term. It is clear that the CD-k is not directly minimizing the likelihood, or equivalently the Kullback Leibler (KL) divergence between the data distribution $p_0(\bm{s})$ and the Boltzmann one $p(\bm{s})$. In reality it minimizes the KL divergence $D_{KL}(p_0||p_k)$  between the data distribution $p_0$ and the distribution obtained after $k$ MC steps $p_k$ that is defined as 
\begin{align*}
    D_{KL}(p_0||p_k) &= \sum_{\{\bm{s}\}} p_0(\bm{s}) \log \frac{p_k(\bm{s})}{p_0(\bm{s})} \\
    p_k(\bm{s}_k) &= \sum_{ \{s_0, \dots, s_{k-1}\}} \sum_{ \{\tau_0, \dots, \tau_{k-1}\}} \left[ \prod_{l=1}^{k} p(\bm{s}_{l}|\bm{\tau}_{l-1})  p(\bm{\tau}_{l-1}|\bm{s}_{l-1}) \right] p_0(\bm{s}_0)
\end{align*} 

\noindent In\ucite{carreira2005contrastive} it is argued that this procedure is roughly equivalent to minimizing the following KL difference 
\[
    \mathcal{L}_{\rm CDk} = D_{KL}(p_0 || p) - D_{KL}(p_k || p),
\]

\noindent up to an extra term considered to be small without much theoretical guaranty. The major drawback of this method is that the phase space of the learned RBM is never explored since we limit ourselves to $k$ MC steps around the data configurations, therefore it can lead to estimate very poorly the probability distribution for configurations that lie ``far away'' from the dataset. A simple modification has been proposed to deal with this issue in~\ucite{tieleman2008training}. The new algorithm is called persistent-CD (pCD) and consists of having again a set of parallel MC chains, but instead of using the dataset as initial condition, they are first initialized from random initial conditions and then the state of the chains is saved from one update of the parameters to the next one. In other words, the chains are initialized one time at the beginning of the learning and are then constantly updated a few MC steps further at each update of the parameters. In that case, it is not longer needed to have as many chains as the number of samples in the mini-batch even though in order to keep the statistical error comparable between the positive and the negative term it should be of the same order.
More details can be found in\ucite{tieleman2008training} about PCD and in\ucite{fischer2014training} for a more general introduction to the learning behavior using MC. In Section~\ref{sec:learning_rbm} we will intend to understand some theoretical and numerical aspect of the RBMs learning process.

\section{Overview of various RBM settings}
\label{sec:link_rbm}

Before investigating the learning behavior of RBMs, let us have a glimpse at various RBM settings and their relation to other models, by looking at common possible priors used for the visible and hidden nodes.

\subsection{Gaussian-Gaussian RBM}
\label{sec:gaussrbm}
The most elementary setting is the linear RBM, where both visible and hidden nodes have Gaussian priors:
\begin{align*}
    q_v(s_i) &= \frac{1}{\sqrt{2 \pi \sigma_v^2}} \exp\left( -\frac{s_i^2}{2 \sigma_v^2}\right) \\
    q_h(\tau_a) &= \frac{1}{\sqrt{2 \pi \sigma_h^2}} \exp\left( -\frac{\tau_a^2}{2 \sigma_h^2}\right)
\end{align*}
\noindent with $\sigma_v$ and $\sigma_h$ the intrinsic variance of the visible and hidden variables respectively.
After summing over hidden variables we get a multi-variate Gaussian distribution over the visible ones. If not very sophisticated, the model is yet interesting because it presents a non-trivial learning dynamics that can be written exactly\ucite{karakida2014analyzing,karakida2016dynamical,decelle2018thermodynamics,decelle2017spectral}.
 When using Gaussian prior, the corresponding activation function $p(\bm{\tau}|\bm{s})$ are Gaussian centered on $\sigma_h^2 \sum_i w_{ia} s_i$:
\[
    p(\bm{\tau}|\bm{s}) \propto \prod_a \exp\left( -\frac{\tau_a^2}{2 \sigma_h^2} + \tau_a \sum_i w_{ia} s_i \right).
\]

Let us write the marginal distribution over the visible nodes $p(\bm{s})$ (we omit the hidden bias since it can be canceled by a redefinition of the visible one), starting from eq. (\ref{eq:bolt}) and integrating over the hidden variables we get
\begin{align}
    p(\bm{s}) &= \frac{1}{Z} \prod_i \left( e^{-\frac{s_i^2}{2 \sigma_v^2} + s_i \theta_i} \right)  \prod_a \left[ \int d\tau_a \exp\left( -\frac{\tau_a^2}{2 \sigma_h^2} + \sum_i s_i w_{ia} \tau_a \right) \right] \nonumber \\
    &= \frac{1}{Z} \prod_i \left( e^{-\frac{s_i^2}{2 \sigma_v^2} + s_i \theta_i} \right) \prod_a \exp\left( \frac{\sigma_h^2}{2} \sum_{ij} s_i w_{ia} w_{ja} s_j \right) \nonumber \\
    &= \frac{1}{Z} \exp\left( - \bm{s}^T \left[ \frac{\bm{1}}{2\sigma_v^2} - \frac{\sigma_h^2}{2} \bm{w} \bm{w}^T \right] \bm{s} + \bm{s}^T \bm{\theta} \right) = \frac{1}{Z} \exp\left( - \bm{s}^T \bm{A} \bm{s} + \bm{s}^T \bm{\theta} \right) \label{eq:gauss:p}
\end{align}
 
\noindent where we define the precision matrix $\bm{A} \equiv \frac{\bm{1}}{2\sigma_v^2} - \frac{\sigma_h^2}{2} \bm{w} \bm{w}^T$. Now we can also identify the conditions for the existence of the measure $p(\bm{s})$. We need the matrix $\bm{A}$ to be strictly positive definite, hence that the highest eigenvalue of $\bm{w} \bm{w}^T$ remains strictly below $1/(\sigma_v^2 \sigma_h^2)$. More interestingly, the Gaussian prior let us write in closed form the stochastic gradients (in fact we solve the deterministic equation, not the stochastic one), hence given us some hints on the nature of the learning dynamics of non-linear RBMs, since in any case we expect a linear regime to take place at the beginning of the learning process. In the present case, we can rewrite eq. (\ref{eq:sgd:w}) as 
\begin{align}
    \frac{\partial \mathcal{L}}{\partial w_{ia}} &= \frac{1}{M} \sum_d s_i^{(d)} \sigma_h^2 \sum_j s_j^{(d)} w_{ja} - \sigma_h^2 \langle s_i \sum_j s_j \rangle w_{ja} \nonumber \\
    &= \sigma_h^2 \left( \sum_j C_{ij}w_{ja} - \sum_j\langle s_i s_j\rangle w_{ja} \right) \nonumber \\
    &= \sigma_h^2 \left( \sum_j C_{ij}w_{ja} - \sum_j A_{ij}^{-1} w_{ja} \right) \label{eq:sgd:gauss}
\end{align}

\noindent where $C_{ij} = \langle s_i s_j \rangle_{\rm data} = M^{-1} \sum_d s_i^{(d)} s_j^{(d)}$ is the correlation between the nodes $i$ and $j$ in the dataset, and $\bm{A}^{-1}$ the inverse of the precision matrix. At this point, following\ucite{decelle2018thermodynamics}, it is convenient to use the singular value decomposition (SVD) of $\bm{w}$. We note $w_{ia} = \sum_\alpha u_i^\alpha w_\alpha v_a^\alpha$ the eigen-decomposition of the rectangular weight matrix $\bm{w}$, where the matrix $\bm{u}$ and $\bm{v}$ correspond to the left (resp. right) eigenvectors of $\bm{w}$ associated to the visible (resp. hidden) variables and $w_\alpha$ the eigenvalue associated to the mode $\alpha$. As can be seen in eq. (\ref{eq:gauss:p}), this transformation will diagonalize the interaction term of the Hamiltonian of the system. We can now make the following change of variables
\begin{equation*}
  \hat{s}_\alpha = \sum_i u_i^\alpha  s_i \;\text{ and }\; \hat{\tau}_\alpha = \sum_a v_a^\alpha \tau_a
\end{equation*}
under this change of variable, the Gaussian measure factorizes where $\sum_{i,j,a} s_i w_{ia} w_{ja} s_j = \sum_\alpha \hat{s}_\alpha w_\alpha^2 \hat{s}_\alpha$ and therefore
\begin{equation*}
  -\bm{s}^T \bm{A} \bm{s} = -\frac{1}{2} \sum_\alpha \hat{s}_\alpha \frac{1-\sigma_v^2 \sigma_h^2 w_\alpha^2}{\sigma_v^2} \hat{s}_\alpha
\end{equation*}
Writing the distribution in this new basis we obtain
\[
    p(\hat{\bm{s}}) \propto \prod_\alpha \exp\left( -\frac{ \hat{s}_\alpha^2}{2 } \frac{1-\sigma_v^2 \sigma_h^2 w_\alpha^2}{\sigma_v^2} \right)
\]
%with $\hat{s}_\alpha$ the components of the visible configuration on $u_\alpha$
\noindent Hence, we can obtain an exact equation for the gradient in the basis of the SVD of the weight matrix $\bm{w}$. First, we project the equation eq. (\ref{eq:sgd:gauss}) on the modes $\alpha-\beta$ of the SVD of $\bm{w}$
\begin{align*}
    \left( \frac{\partial \mathcal{L}}{\partial \bm{w}} \right)_{\alpha \beta} &= \sum_{ia} u_i^\alpha \frac{\partial \mathcal{L}}{\partial w_{ia}} v_a^\beta =  \sum_{ia} u_i^\alpha \left[ \langle s_i \tau_a \rangle_{\rm data} - \langle s_i \tau_a \rangle_{\mathcal{H}} \right] v_a^\beta \\
    &= \langle \hat{s}_\alpha \hat{\tau}_\beta \rangle_{\rm data} - \langle \hat{s}_\alpha \hat{\tau}_\beta \rangle_{\mathcal{H}} 
     % &= \left( \frac{d \bm{w}}{dt} \right)_{\alpha \beta}
\end{align*}

\noindent Now to simplify we discard the fluctuations associated to the stochastic gradient, by considering instead the full gradient and an infinitesimal learning rate such that we can consider the iteration time to be continuous and identify $\frac{\partial \mathcal{L}}{\partial w_{ia}} \sim \frac{dw_{ia}}{dt}$. As a result we obtain the time derivative of the matrix $\bm{w}$ decomposed over its eigenmodes
\begin{align*}
    \left( \frac{d \bm{w}}{dt} \right)_{\alpha \beta} &= \sum_{ia} u_i^\alpha\left( \frac{d}{dt} \sum_\gamma u_i^\gamma w_\gamma v_a^\gamma \right) v_a^\beta \\
    &= \sum_{ia\gamma} u_i^\alpha u_i^{\gamma} \frac{dw_{\gamma}}{dt} v_a^\gamma v_a^\beta + u_i^\alpha \frac{du_i^\gamma}{dt} w_\gamma v_a^\gamma v_a^\beta + u_i^\alpha u_i^\gamma  w_\gamma \frac{dv_a^\gamma}{dt} v_a^\beta \\
    &= \delta_{\alpha \beta} \frac{dw_{\alpha}}{dt} + (1-\delta_{\alpha \beta})\left( \bm{u}^\alpha \frac{d\bm{u}^\beta}{dt} w_\alpha + w_\beta \frac{d\bm{v}^\alpha}{dt}\bm{v}^\beta \right)
\end{align*}

\noindent This equation shows that, the gradient update of $\bm{w}$ can be decomposed when projected on the SVD basis of $\bm{w}$ into a gradient over the mode $w_\alpha$ and a rotation of the matrices $\bm{u}^\alpha$ and $\bm{v}^\alpha$. Noticing first that $\langle \hat{s}_\alpha \hat{\tau}_\alpha \rangle = \sigma_h^2 w_\alpha \langle \hat{s}^2_\alpha \rangle$, we therefore end up with the following dynamics for the singular values $w_\alpha$:
\begin{align}
    \frac{dw_\alpha}{dt} &= \left( \frac{d\bm{w}}{dt} \right)_{\alpha \alpha} = \langle \hat{s}_\alpha \hat{\tau}_\alpha \rangle_{\rm data} - \langle \hat{s}_\alpha \hat{\tau}_\alpha \rangle_{\mathcal{H}} \nonumber\\
    &= \sigma_h^2 w_\alpha \left( \langle \hat{s}^2_\alpha \rangle_{\rm data} -\langle \hat{s}^2_\alpha \rangle_{\mathcal{H}}  \right) \nonumber\\
    &= \sigma_h^2 w_\alpha \left( \langle \hat{s}^2_\alpha \rangle_{\rm data} - \frac{\sigma_v^2}{1-\sigma_v^2 \sigma_h^2 w_\alpha^2}  \right) \label{eq:sgd:walpha}
\end{align}

\noindent where in eq (\ref{eq:sgd:walpha}) $\langle \hat{s}^2_\alpha \rangle_{\rm data}$ denotes the variance of the components of the data on the mode $\alpha$:
\[
   \langle \hat{s}^2_\alpha \rangle_{\rm data} = \sum_{ia} u_i^{\alpha} \left( \frac{1}{M} \sum_{ij} s_i^{(d)} s_j^{(d)} \right) u_j^{\alpha}
\]

\noindent This first result tells us that when keeping the matrices $\bm{u}$ and $\bm{v}$ fixed, the SGD on the mode $w_\alpha$ will adjust the value of $w_\alpha$ such that the r.h.s matches the variance in the direction given by $\bm{u}^\alpha$, giving the following limit values: 
\begin{equation}
    w_\alpha^2 = \left\{ 
      \begin{array}{ll}
        \frac{ \langle \hat{s}^2_\alpha \rangle_{\rm data} -\sigma_v^2}{\sigma_v^2 \sigma_h^2  \langle \hat{s}^2_\alpha \rangle_{\rm data} } & \text{ if }  \langle \hat{s}^2_\alpha \rangle_{\rm data}  > \sigma_v^2\\
        0 & \text{ if }  \langle \hat{s}^2_\alpha \rangle_{\rm data}  < \sigma_v^2
      \end{array}
               \right.
   \label{eq:w_max}
\end{equation}

\noindent We remark that, if the empirical variance given by the data is smaller than the prior variance of the visible variables the corresponding mode is filtered out. The evolution of the matrices $\bm{u}^\alpha$ and $\bm{v}^\alpha$ can also be obtained~\ucite{decelle2017spectral} from the following expression in the present case\footnote{Actually these equations are given with a wrong sign in~\ucite{decelle2017spectral} which is corrected here.}
\begin{align}
     \Omega_{\alpha \beta}^{u} &\equiv \left( \frac{d \bm{u}^\alpha}{dt}\right)^T \bm{u}^\beta =-(1-\delta_{\alpha \beta})\sigma_h^2\left(\frac{w_\beta - w_\alpha}{w_\alpha + w_\beta} - \frac{w_\beta + w_\alpha}{w_\alpha - w_\beta} \right) \langle s_\alpha s_\beta \rangle_{\rm data} \label{eq:sgd_u} \\
    \Omega_{\alpha \beta}^{v} &\equiv \left( \frac{d \bm{v}^\alpha}{dt}\right)^T \bm{v}^\beta = -(1-\delta_{\alpha \beta})\sigma_h^2\left(\frac{w_\beta - w_\alpha}{w_\alpha + w_\beta} + \frac{w_\beta + w_\alpha}{w_\alpha - w_\beta} \right) \langle s_\alpha s_\beta \rangle_{\rm data} \label{eq:sgd_v} 
\end{align}

\noindent of the infinitesimal rotations of the vectors $\bm{u}^\alpha$ and $\bm{v}^\alpha$. In the particular case of the Gaussian-Gaussian RBM, we can note the absence of term averaged over the model $\langle . \rangle_{\mathcal{H}}$. This is due to the fact that the SVD corresponds to the eigendecomposition of the RBM measure (that is, the Gaussian measure factorizes over the singular modes) and that the eqs (\ref{eq:sgd_u}-\ref{eq:sgd_v}) involve correlation between modes $\alpha \neq \beta$ which are zero here. From eq. (\ref{eq:sgd_u}-\ref{eq:sgd_v}), we see that a steady state is found when a direction $\bm{u}^\alpha$ is found that diagonalizes the empirical covariance matrix of the dataset.

In short, the Gaussian-Gaussian RBM learns the principal components of the dataset and for each principal axes the weight matrix is adjusted until the strength of the corresponding modes $w_\alpha$ reaches the value given by eq. (\ref{eq:w_max}). Of course, modes above threshold acquire a variance which matches the variance of the dataset in this direction $ \langle s_\alpha^2 \rangle_{\mathcal{H}} = \langle s_\alpha^2 \rangle_{\rm data}$. We can somehow say that the Gaussian-Gaussian RBM is performing a sort of SVD of the dataset, keeping only the modes above a given threshold. It is worth noting that an analysis has been done in\ucite{karakida2016dynamical} where it is shown that updating the parameters of the model using the $k$CD approximation converges toward the same solution as the one obtained by maximizing the likelihood of the model.

\begin{figure}[ht]
\centering{
\resizebox*{0.3\textwidth}{!}{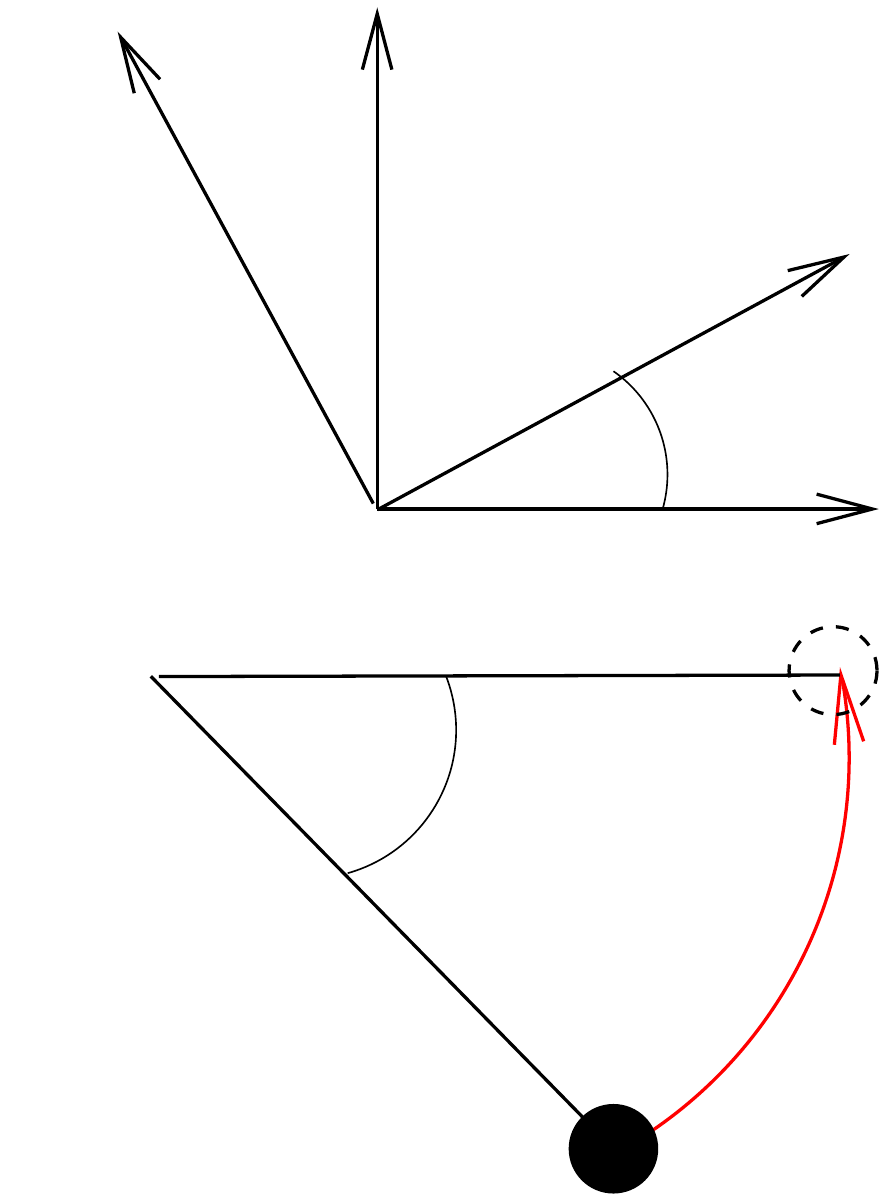}
\hspace{0.5cm}\includegraphics[width=0.62\textwidth]{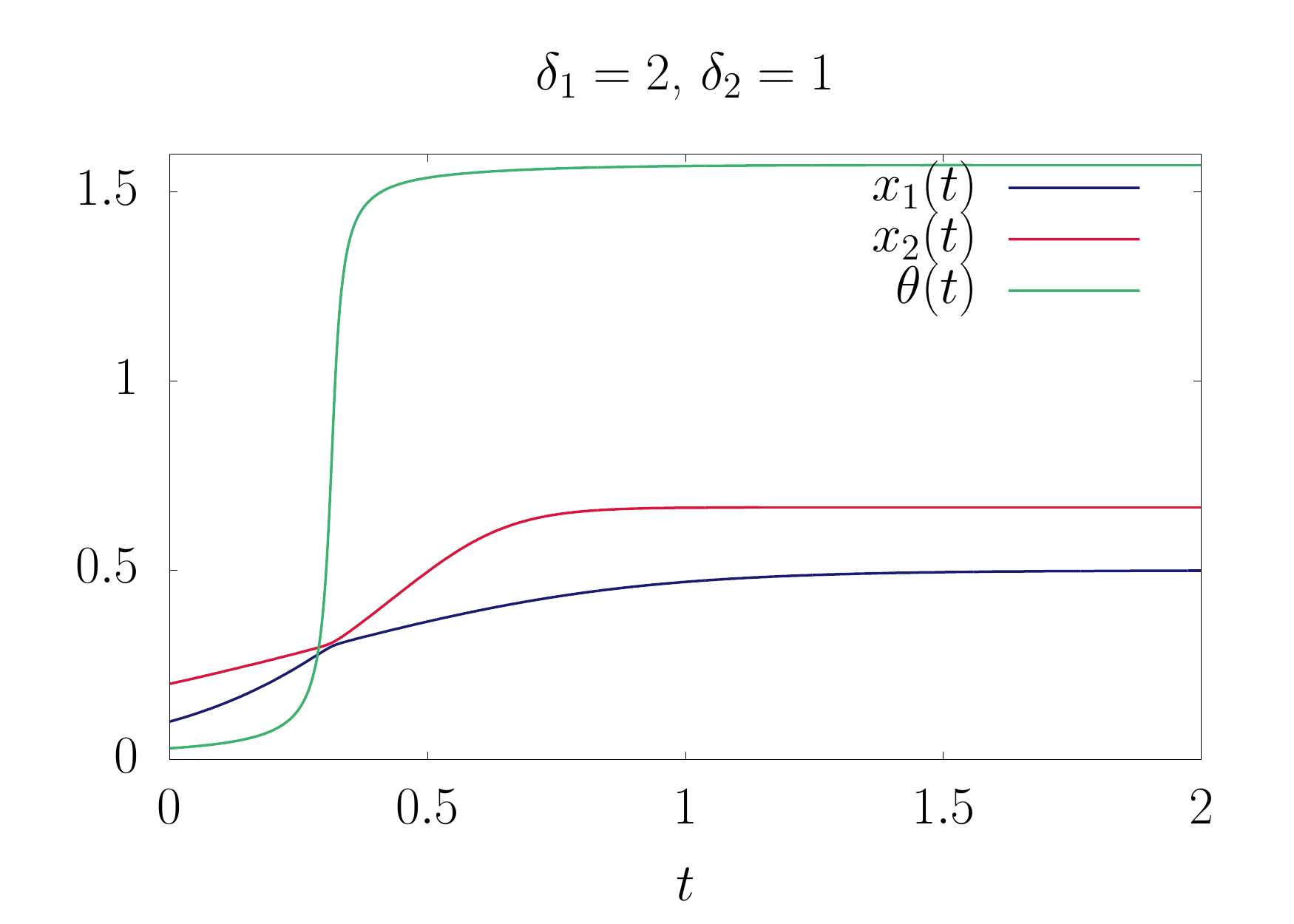}}
\caption{Angle between the reference basis given by the data and the moving one given by the RBM shown on the up left panel. Equivalence with the motion of a pendulum is indicated on the left bottom panel.
 Solution of (\ref{eq:x1},\ref{eq:x2},\ref{eq:theta}) of two coupled modes in the linear RBM (right panel)}\label{fig:pendulum}
\end{figure}

We can illustrate the learning mechanism in simple cases where it is possible to solve explicitly the dynamics. First assume that the  RBM has found the principal axes, i.e.  consider the matrices $\bm{u}$ and $\bm{v}$ to be fixed. In this case the quantity $\langle \hat s_\alpha^2 \rangle_{\rm data}$ remains constant. Letting
\[
x_\alpha = \sigma_v^2\sigma_h^2 w_\alpha^2\qquad\text{and}\qquad \delta_\alpha = \frac{\langle \hat s_\alpha^2\rangle_{\rm Data}-\sigma_v^2}{\sigma_v^2},
\]
and rescaling time as $t\sigma_v^2\sigma_h^2\to t$, equation~(\ref{eq:sgd:walpha}) then rewrites as
\[
\dot x_\alpha = 2x_\alpha\bigl(\delta_\alpha-\frac{x_\alpha}{1-x_\alpha}\bigr)
\]
and we obtain a solution of the form
\[
x_\alpha(t) = f_\alpha^{-1}(\delta_\alpha t),
\]
with
\[
f_\alpha(x) = \log\frac{x}{x_\alpha(0)}-\frac{1}{1+\delta_\alpha}\log\frac{\gamma_\alpha-x}{\gamma_\alpha-x_\alpha(0)},\qquad\text{and}\qquad
\gamma_\alpha = \frac{\delta_\alpha}{1+\delta_\alpha}.
\]
For $\delta_\alpha\ll 1$ we get a sigmoid type behaviour
\[
\frac{x_\alpha(t)}{x_\alpha(0)}  = \frac{\delta_\alpha e^{\delta_\alpha t}}{\delta_\alpha+x_\alpha(0)\bigl(e^{\delta_\alpha t}-1)}.
\]
To illustrate the rotation of the modes, consider now the situation where there are $2$ modes $u_\alpha$, $\alpha=1,2$ which are a linear combination of
two dominant modes of the data $\{\hat u_1,\hat u_2\}$ with identical orientation taken in this order, all other modes considered to be already properly aligned with the data. Let then $\theta$ represent the angle between $u_1$ and $\hat u_1$ (and also between $u_2$ and $\hat u_2$ see Figure~(\ref{fig:pendulum}). Equation~(\ref{eq:sgd_u}) for this pair of modes rewrites then as
\[
\frac{d\theta}{dt} = -\sigma_h^2\Bigl(\frac{w_\alpha^2+w_\beta^2}{w_\alpha^2-w_\beta^2}\Bigr)\langle s_1 s_2\rangle_{\rm Data}(t),
\]
with
\begin{align*}  
  \langle s_1 s_2\rangle_{\rm Data}(t) &= \cos\theta\sin\theta\bigl(\langle s_2^2\rangle_{\rm Data}-\langle s_1^2\rangle_{\rm Data}\bigr),\\[0.2cm]
  \langle s_1^2\rangle_{\rm Data}(t) &= \cos^2\theta\langle s_1^2\rangle_{\rm Data}+\sin^2\theta\langle s_2^2\rangle_{\rm Data}\\[0.2cm]
  \langle s_2^2\rangle_{\rm Data}(t) &= \sin^2\theta\langle s_1^2\rangle_{\rm Data}+\cos^2\theta\langle s_2^2\rangle_{\rm Data}
\end{align*}
so that finally we get a dynamical system of the form
\begin{align}
\dot x_1 &= 2x_1\bigl(\delta_1\cos^2\theta+\delta_2\sin^2\theta-\frac{x_1}{1-x_1}\bigr) \label{eq:x1}\\[0.2cm]
\dot x_2 &= 2x_2\bigl(\delta_1\sin^2\theta+\delta_2\cos^2\theta-\frac{x_2}{1-x_2}\bigr) \label{eq:x2}\\[0.2cm]
\dot\theta &= -\frac{1}{2}(\delta_1-\delta_2)\frac{x_1+x_2}{x_1-x_2}\sin(2\theta) \label{eq:theta}
\end{align}
Note that at fixed $x_1$ and $x_2$ the dynamics of $\theta$ corresponds to the motion of a pendulum w.r.t the variable $\theta' = 4\theta$ shown on Figure~\ref{fig:pendulum}.

\subsection{Gaussian-Spherical}
\label{sec:sphrbm}
The Gaussian-Gaussian case is interesting as a solvable model of RBM but of limited scope, since $p(\bm{s})$ reduces in the end to a multivariate Gaussian. Next, a simple non-linear RBM which remains exactly solvable is based on the so-called spherical model\ucite{berlin1952spherical,stanley1968spherical}. For this model, it is possible to compute the phase diagram and the equilibrium states once the coupling matrix is given ---more precisely, when the spectral density of the coupling matrix is given. Here we chose the following priors to impose a spherical constraint on the hidden nodes:
\begin{align*}
    q_v(s_i) &= \frac{1}{\sqrt{2 \pi \sigma_v^2}} \exp\left( -\frac{s_i^2}{2 \sigma_v^2}\right) \\
    q_h(\bm{\tau}) &= \delta\left(\sum_a \tau_a^2 - \bar{\sigma} \sqrt{N_h N_v}\right)
\end{align*}

\noindent where $\bar{\sigma}$ is a parameter of the model\ucite{decelle2020gaussian}. The interest of such an  RBM is first that the spherical constraint can be dealt with analytically\ucite{decelle2020gaussian,genovese2020legendre}. Secondly the model can exhibit a phase transition unlike the Gaussian-Gaussian case. Absorbing the parameter $\sigma_v^2$ in the definition of the weight matrix, to follow the computation of\ucite{decelle2020gaussian}, a simple analysis in the thermodynamic limit tells us that the phase transition takes place when the parameter $\bar{\sigma}$ exceeds the value $\sigma_c$, where $\sigma_c$ depends on the value of the highest mode $w_{\rm max}$ and of the form of the spectrum of $\bm{w}$ (typically, $\sigma_c^2 \propto 1/w_{\rm max}^2$, where the pre-factor depends on the form of the spectrum). The condensation along this mode of the visible (resp. hidden) magnetization is then given by
\begin{align*}
    m_\alpha &= \frac{1}{\sqrt{L}}\sum_i u_i^\alpha \langle s_i \rangle_{\mathcal{H}} = w_{\rm max}\bar{\sigma}\sqrt{\bar{\sigma}^2 - \bar{\sigma}_c^2 } \\
    \bar{m}_\alpha^2 & =\frac{1}{\sqrt{L}}\sum_a v_a^\alpha \langle \tau_a \rangle_{\mathcal{H}} = \sqrt{\bar{\sigma}^2 - \bar{\sigma}_c^2 }
\end{align*}

\noindent \noindent where we defined $L=\sqrt{N_v N_h}$. This  type of RBM is again of limited scope to represent data. In the thermodynamic limit a finite number $n={\cal O}(1)$ of modes can condense. They necessarily accumulate at the top of the spectrum of the weight matrix and represent a distribution concentrated on an $n$-dimensional sphere in absence of external fields while other non-condensed modes are responsible for transverse Gaussian fluctuations. The dynamical aspect of this model will be discussed in Section~(\ref{sec:learning_rbm}).

\noindent To end up this section let us also  mention that the finite size regime is amenable to an exact analysis when restricting the weight matrix spectrum to have the property of being doubly degenerated (see\ucite{decelle2020gaussian} for details).

\subsection{Gaussian-Softmax}
The case of the Gaussian mixture if rarely viewed like that, fits actually perfectly the RBM architecture. Consider here the case of Gaussian visible nodes and a set of discrete $\{0,1\}$ hidden variables with a constraint corresponding to the softmax activation function~\ucite{nijman1997symmetry}:
\begin{align*}
    q_v(s_i) &= \frac{1}{\sqrt{2 \pi \sigma_v^2}} \exp\left( -\frac{s_i^2}{2 \sigma_v^2}\right), \\
    q_h(\bm{\tau}) &= \prod_a \left(\delta_{\tau_a,0} + \delta_{\tau_a,1} \right)\delta_{\sum_{b} \tau_{b},1}.
\end{align*}

\noindent With this formulation, we indeed see that the conditional probability of activating a hidden node is a softmax function
\begin{equation*}
    p(\tau_a=1|\bm{s}) = \frac{\exp\left( \sum_i w_{ia} s_i  + \eta_a\right)}{\sum_b \exp\left( \sum_i w_{ib} s_i + \eta_b\right)}.
\end{equation*}

\noindent It is easy from this expression to recognize the equations of the Gaussian mixture model (GMM)\ucite{mackay2003information,bishop2006pattern}, where the latent variable $\tau_a$ indicates if a sample belong or not to the center $a$. The position of the associated center is given by the vector $\bm{w}_a$. It is even clearer when writing the marginal over the visible nodes after summing over the hidden nodes in eq. (\ref{eq:bolt})
\begin{align*}
    p(\bm{s}) &= \frac{1}{Z} \sum_a \exp\left(\eta_a+\sum_i -\frac{s_i^2}{2\sigma_v^2} + \theta_i s_i + s_i w_{i a}  \right) \\
    &= \frac{1}{Z} \sum_a \exp\left(\eta_a+\sum_i -\frac{1}{2\sigma_v^2}\left(s_i - \sigma_v^2[w_{ia} + \theta_i]^2 \right)^2  +\frac{1}{2}\sigma_v^2 [w_{ia} + \theta_i]\right)\\
    &=  \frac{1}{Z'} \sum_a \rho_a \exp\left(\sum_i -\frac{1}{2\sigma_v^2}\left(s_i - \sigma_v^2[w_{ia} + \theta_i]^2 \right)^2 \right)
\end{align*}

\noindent by identifying 
\begin{equation}
  \rho_a \equiv \frac{\exp\left(\eta_a + \sum_i \left(\sigma_v^2 w_{ia} + \sigma_v^2 \theta_i \right)^2 \right)}{\sum_b \exp\left(\eta_b + \sum_i \left(\sigma_v^2 w_{ib} + \sigma_v^2 \theta_i \right)^2 \right)} \label{eq:def:rho}
\end{equation}
\noindent the weight of the mode $a$ in the Gaussian mixture centered in $\bm{w}_a$. Now we can see that the extra parameter $\theta_i$ can be absorbed in the definition of the weight matrix $w'_{ia} = w_{ia} + \theta_i$. It turns out that the positive term of the gradient in equation (\ref{eq:sgd:w}) (ignoring $\rho_a$) corresponds to the gradient that is obtained in the GMM. This can be reformulated into the Expectation Maximization (EM) update by considering that $p(\tau_a|\bm{s})$ do not depend on $w_{ia}$, hence doing the ``expectation'' step:
\begin{equation}
    \langle s_i \tau_a \rangle_{\rm data} = \frac{1}{M} \sum_d \left(s_i^{(d)} -  \sigma_v^2 w_{ia}\right) p(\tau_a | \bm{s}^{(d)}) \label{eq:gmm:em1}
\end{equation}

\noindent If we impose that the gradient is zero, doing now the ``maximization'' step, we obtain
\begin{equation}
    w_{ia}^{(t+1)} = \frac{\sum_d s_i^{(d)} p(\tau_a|\bm{s}^{(d)})}{\sigma_v^2 \sum_d p(\tau_a|\bm{s}^{(d)})} \label{eq:gmm:em2}
\end{equation}

\noindent where the l.h.s. is to be understood as the new values for the parameters $w_{ia}$ while the conditional distribution on the r.h.s. depends on $w_{ia}^{(t)}$. For an RBM, one would also compute the negative term of the gradient, involving the derivative of $\rho_a$ w.r.t. $w_{ia}$. We obtain the negative term
\begin{equation}
  \langle s_i \tau_a \rangle_{\mathcal{H}} =  \frac{1}{M} \sum_d  \sigma_v^2 w_{ia}\left[p(\tau_a | \bm{s}^{(d)}) - \rho_a \right] \label{eq:rho:gmm}
\end{equation}

\noindent Again, we can recover with equation (\ref{eq:rho:gmm}) the EM update for the density of the Gaussian mode $a$ in the GMM, by first considering that the conditioned distribution $p(\tau_a|\bm{s}^{(d)})$ does not depend on $w_{ia}$ (expectation step) and by putting the l.h.s. to zero (maximization step). The fact that when using the RBM formalism we do not obtain directly the same EM equations as in the GMM is due to the different parametrization of the parameters. In the GMM, the density of each Gaussian is defined right from the beginning as an independent parameter while when using the RBM, the density of the Gaussian depends on other parameters such as the weight matrix $\bm{w}$.

\paragraph{Phase transition in the learning process ---} an interesting phenomena occurs in this model when learning position of the centers of the Gaussian while submitting the variances $\sigma_v$ of the Gaussian to an annealing process\ucite{rose1990statistical}. First of all, starting from a very high variance (equivalently, very high temperature), we can convince ourselves that the learning will end up finding the center of mass of the dataset. Let us therefore consider that we centered the dataset beforehand: $\sum_d s_i^{(d)} = 0$, $\forall i$. Then, reducing slowly the variance of each component of the mixture, we can look for the moment at which point the degenerate solution corresponding to all the centers placed at the center of masses of the dataset  becomes unstable. Linearizing the EM equations (\ref{eq:gmm:em2}) around this point with $\eta_a=0$ and $w_{ia} \approx 0 + \epsilon_{ia}$, where the $\bm{\epsilon}$ are small perturbations, we can derive the threshold where the linear perturbations get amplified. The linear stability analysis leads to the following equations for the perturbation $\epsilon$:
\begin{align*}
    \epsilon_{ia}^{(t+1)} &\approx \frac{\sum_d s_i^{(d)} (1 + \sum_j s_j^{(d)} \epsilon_{ja}^{(t)} - \frac{1}{N_h}\sum_{jb} s_j^{(d)}\epsilon_{jb}^{(t)})}{\sigma_v^2 \sum_d (1 + \sum_j s_j^{(d)}\epsilon_{ja}^{(t)} - \frac{1}{N_h}\sum_{jb} s_j^{(d)} \epsilon_{jb}^{(t)})} \\
    &= \frac{1}{\sigma_v^2}\sum_j c_{ij} \left( \epsilon_{ja}^{(t)} - \frac{1}{N_h} \sum_b \epsilon_{jb}^{(t)} \right)
\end{align*}

\noindent where $c_{ij}$ is the covariance matrix of the dataset. From this expression, one sees that when the variance is higher that the largest eigenvalue $\Lambda_C$ of $\bm{c}$, i.e. $\sigma_v^2> \Lambda_C$, the solution $w_{ia}=0$ is stable. Then, when $\sigma_v^2 < \Lambda_C$, the solution is unstable and the system starts to learn something more about the dataset besides its center of mass. It is interesting to note that this threshold is very similar to the one obtained in equation (\ref{eq:w_max}) for the Gaussian-Gaussian RBM. In this model, it is then possible to study the cascade of phase transition, occurring in a hierarchical way on structured datasets\ucite{Kloppenburg1997,Kappen2000}. We stress here that, even if it is possible to project the learning equations on the SVD of the weight matrix as in the two previous analysis, it does not provide much more insight since this case cannot be solved exactly by this transformation.

It is also interesting to investigate the behavior of the exact gradient (not using EM) in the presence of a learning rate $\gamma$. When using the gradient, the update equations are given by $w_{ia}^{(t+1)} = w_{ia}^{(t)} + \gamma \Delta w_{ia}$. In that case we obtain the following equation for the linear stability
\begin{equation*}
    \epsilon_{ia}^{(t+1)} = (1-\gamma) \epsilon_{ia}^{(t)} + \frac{\gamma}{\sigma_v^2} \sum_j c_{ij} \left( \epsilon_{ja}^{(t)} - \frac{1}{N_h} \sum_b \epsilon_{jb}^{(t)} \right)
\end{equation*}

\noindent Interestingly, the threshold does not depend on the value of $\gamma$ in that case, meaning that the instability is a generic properties of the learning dynamics. The only change is the speed with which the instabilities will develop.

\subsection{Bernoulli-Gaussian RBM}

The next case is the Bernoulli-Gaussian RBM where we consider the following prior
\begin{align*}
    q_v(s_i) &= \frac{1}{2}\left( \delta_{s_i,0} + \delta_{s_i,1}\right), \\
    q_h(\tau_a) &= \frac{1}{\sqrt{2 \pi \sigma_h^2}} \exp\left( -\frac{\tau_a^2}{2 \sigma_h^2}\right).
\end{align*}
\noindent Again, a Gaussian prior implies that the activation function is Gaussian. It is interesting to consider this version of the RBM through its relation with the Hopfield model\ucite{hopfield1982neural} realized in\ucite{BaBeSaCo}. Since the hidden variables are  Gaussian they can be integrated out which leads to a simple analytical form for the marginals of the visible variables. In some recent works, the opposite approach has been done, starting with an Hopfield model and expressing it as a RBM using the Hubbard-Stratonovitch (HS) transformation (expressing the exponential of a square as Gaussian integral) to decouple the interactions between each spin\ucite{mezard2017mean,shimagaki2019selection}. After integrating over the hidden nodes in eq. (\ref{eq:bolt}), we end up with the following distribution
\begin{equation*}
    p(\bm{s}) = \frac{1}{Z}\exp\left( \frac{\sigma_h^2}{2}\sum_{ij} s_i s_j \left[\sum_a w_{ia} w_{ja} \right] \right)
\end{equation*}

\noindent We recognize a Hopfield model where the patterns are given by the weights $w_{ia}$ of the RBM and the effective coupling between two variables $i$ and $j$ is $J_{ij} = \sum_a w_{ia} w_{ja}$. We can also consider that the variances of the hidden nodes is related to the temperature of the model. 

Some experiments have been conducted in\ucite{decelle2019inverse} in order to compare the learning process of the Hopfield model versus the Bernoulli-Gaussian RBM on artificial data generated from an Hopfield model with dicrete patterns. It is interesting to note that, when assuming discrete patterns, the inverse procedure can be formulated in terms of an approximated Hopfield model. Thus, the inference of the pattern can be done directly using a set of TAP equations of the Hopfield model, and it has been shown that the artificial patterns were inferred exacltly. When using the RBM's formulation, in the absence of information over the patterns, only the subspace covered by the patterns were retrieved with a weak overlap with the true patterns. In fact, in that case the marginal over the visible nodes is a function of $\bm{w} \bm{w}^T$, which is invariant by rotation of the $\bm{v}$ matrix. It explains why the learned weight matrix in the RBM context does not overlap with the true patterns.

With this machine, it is also possible to impose a maximum rank in order to reduce the number of parameters needed to describe the dataset giving the possibility of a trade-off between a good description of the dataset and the number of parameters. This properties has been used in\ucite{shimagaki2019selection} to find global patterns in protein foldings, using the RBM version of the Hopfield model with $q$ discrete states.

\subsection{Gaussian-Bernoulli RBM}

At this point we now focus on models where the hidden layers will have a stronger impact. The integration of the hidden layer will not end up in a simple analytical form and therefore will make it difficult to understand the effect of the features and to characterize properly the learning dynamics. We first mention the Gaussian-Bernoulli case dealing with the following priors:
\begin{align}
    q_v(s_i) &= \frac{1}{\sqrt{2 \pi \sigma_v^2}} \exp\left( -\frac{s_i^2}{2 \sigma_v^2}\right), \label{eq:prior:GB1}\\
    q_h(\tau_a) &= \frac{1}{2}\left( \delta_{\tau_a,0} + \delta_{\tau_a,1}\right).  \label{eq:prior:GB2}
\end{align}

\noindent When using the discrete $\{0,1\}$ variables, we obtain the sigmoid activation function for the hidden nodes
\begin{equation*}
    p(\tau_a = 1|\bm{s}) = \frac{1}{1+\exp(-\sum_i w_{ia} s_i + \eta_a)}.
\end{equation*}

\noindent With this parameterization, it is natural to interpret a hidden node $\tau$ as an active feature when $\tau=1$  and an inactive one if $\tau=0$. When responding to a given input through the conditional probability $p(\bm{\tau}|\bm{s})$, the machine is turning on the hidden nodes corresponding to overlapping features with the input. Therefore, the input undergoes a non-linear decomposition on the learned features. Saying it that way, it is somewhat reminiscent of the Independent Component Analysis\ucite{hyvarinen2000independent} where a matrix $X$ is factorized on a set of independent sources or components $\bm{y}$: $\bm{x} = \bm{A}\bm{y}$. The sources here are independent in the sense that they are independently distributed. In the context of ICA, the goal is to find the inverse of the mixing matrix in order to recover the sources from the received signal. Concerning this particular RBM, it is proven\ucite{karakida2016dynamical} that under some assumptions ---  (i) having the same number of visible and hidden nodes, (ii) that the signal comes from a set of independent sources and (iii) that the variance of the visible variables is much smaller than the mean of the signal---  there exists a stable solution for the learning dynamics where the learned weight matrix corresponds to the un-mixing matrix of the signal. In this regime, the RBM acts as an ICA. In other words, if the signal $\bm{s}^d$ used as an input for the RBM can be written as a mixture of sources: $\bm{s} = \bm{A} \bm{y}$, a stable solution of the learning process consists in recovering the inverse mixing matrix in the weight matrix: $\bm{w} = \bm{A}^{-1}$.

To end up with this variant of the RBM, it is interesting to note that the prior variance of the visible variables here is in principle a fixed parameter. It has been noted that when using the prior (\ref{eq:prior:GB1}-\ref{eq:prior:GB2}) the mean of the conditional distribution over the visible $p(s_i|\bm{\tau})$ is stretched by the variance $\sigma_v$. It might be useful to remove this effect by renormalizing the weight matrix and the visible biases as in\ucite{cho2011improved} : $\bm{w} \rightarrow \bm{w}/\sigma_v^2$ and $\theta_i \rightarrow \theta_i/\sigma_v^2$. Using this parametrization, we obtain 
\begin{equation*}
    p(s_i|\bm{\tau}) \approx \mathcal{N}(\theta_i + \sum_a w_{ia} \tau_a,\sigma_v^2),
\end{equation*}

\noindent where $\mathcal{N}$ represents the normal distribution. Note that it is possible to include the learning of these parameters in the likelihood ascent as in\ucite{cho2011improved}. It is however important to stress here that even if appealing, the possibility to tune the variance of each visible node doesn't solve the problem of learning individual variances of separated  clusters in a dataset. Indeed, consider the problem where the dataset is formed of many well-separated clusters with distinct variances. For a given visible node $i$, its variance computed over the whole dataset or instead over a given cluster have no reason to coincide. And the the prior variance if properly learned will only account for the global variance of this node. This should involve  
a more complex setting of the RBM which we won't discuss here in order to account for individual variances of clusters in a complex dataset.

\subsection{Bernoulli-Bernoulli RBM}

The last model here is traditionally the one which is implied when speaking of RBM. In that case both the visible and hidden nodes are in $\{0,1\}$ with the following priors
\begin{align*}
    q_v(s_i) &= \frac{1}{2}\left( \delta_{s_i,0} + \delta_{s_i,1}\right), \\
    q_h(\tau_a) &= \frac{1}{2}\left( \delta_{\tau_a,0} + \delta_{\tau_a,1}\right). 
\end{align*}

\noindent The activation functions are sigmoid functions, for both the hidden and visible nodes
\begin{align}
    p(s_i = 1|\bm{\tau}) &=  \frac{1}{1+\exp(-\sum_a w_{ia} \tau_a + \theta_i)}, \\
    p(\tau_a=1|\bm{s}) &=  \frac{1}{1+\exp(-\sum_i w_{ia} s_i + \eta_a)}. 
\end{align}
In that case, the prior distribution has the advantage of not having any free parameter to be determined. In practice this model is used when dealing with a discrete dataset while the Gaussian-Benoulli is for continuous ones. This model can also be generalized to the case where the hidden nodes take more than two states, see\ucite{yokoyama2019restricted} for more details on this approach.

\paragraph{Rectified Linear Units (RELU) ---} Let us briefly mention how the Bernoulli prior on the hidden nodes can be linked to the RELU activation function\ucite{hahnloser2000digital} for the RBM. In a work by Teh et al\ucite{teh2001rate}, was highlighted one important shortcoming with Bernoulli prior. With the hidden variable in $\{0,1\}$, a given pattern can be expressed  $\tau=1$, or not  $\tau=0$. Therefore the influence of a feature is binary, either $0$, either a fixed amount given by the value of $\bm{w}$: it is not possible to tune this amount as a function of how strongly a hidden node responds to a visible configuration. Of course it is possible for the machine to learn many times the same pattern, but this doesn't seem very efficient. A simple idea to correct this problem is to duplicate many time a hidden node, keeping the same features and bias values. Then, if the probability of turning on this hidden node is $p$, the average number of activated hidden nodes for this feature will be $Np$ giving the possibility to tune the intensity of the feature.

Generalizing this idea, it is possible to construct an infinite number of replica\ucite{nair2010rectified}, adjusting the bias for each of them such that in order to activate more and more neurons it is necessary that the signal $\sum_i w_i s_i$ is stronger and stronger.  Let us focus for a moment on a single hidden nodes with a feature $w_i$ and a bias $\eta$ along with its replicas $a'=1, \dots,N_r$. We denote $r=\sum_i w_i s_i + \eta$ the potential associated for this neuron given the signal $\bm{s}$. The number of activated replica will be given by
\begin{equation}
    \frac{1}{\sqrt{N_r}}\sum_{a'=0}^{N_r-1} {\rm sig}\left(r(1 - a'/\sqrt{N_r} \right)) \underset{N_r \to \infty}{\approx} \log(1+\exp(r)) \label{eq:relu1}
\end{equation}
\noindent where we defined the sigmoid function ${\rm sig}(x)=(1+\exp(-x))^{-1}$. The r.h.s. of eq. (\ref{eq:relu1}) is very close to the RELU activation function ${\rm RELU}(x) = {\rm max}(0,x)$, hence showing that having all these replicas gives a similar activation function as RELU. In practice, it is not very efficient to have a large number of sigmoids for the training algorithm. An approximation is found by using the truncated Gaussian distribution. The average number of activated replica is then given by 
\begin{equation}
    \tau_a = {\rm max}(0,r+ \mathcal{N}(0,\sigma_a)) \label{eq:relu_act}
\end{equation}

\noindent where now $\tau_a$ is a RELU hidden node and $\sigma_a$ is the variance associated with the number of activated replicas for the hidden node $a$. Eq. (\ref{eq:relu_act}) can now be seen as an approximation of the Truncated-Gaussian prior for the hidden nodes
\begin{equation}
    q_h(\tau_a) \propto \delta_{\tau_a>0}\ \exp(-\frac{\tau_a^2}{2 \sigma_h})
\end{equation}

In the following section, we will focus mainly on the Bernoulli-Bernoulli setting, its equilibrium phase diagram and its learning dynamics in the mean-field regime.

\section{Phase diagram of the Bernoulli-Bernoulli RBM}
\label{sec:phase_diag}

In this section, we discuss various aspects of the phase diagram of the Bernoulli-Bernoulli RBM. In the rest of the section we will use $\{\pm1\}$ instead of the usual $\{0,1\}$ for commodity. There are (at least) two series of works dealing with the RBM in the thermodynamic limit, each of them making different hypothesis on the statistical ensemble from which the RBM is taken. In the first one\ucite{barra2018phase,tubiana2017emergence} the weight matrix is taken from a simple statistical ensemble with iid elements and possibly additional sparse constraints on the patterns as will be explained in Section~\ref{sec:rbm:rbm}.  In the second one\ucite{decelle2017spectral,decelle2018thermodynamics} it is assumes that the weight matrix contains a structured part of rank $K={\cal O}(1)$ in addition to a random matrix corresponding to noise; the main results of this approach will be exposed in Section~\ref{sec:svd:rbm}. Both approaches are based on the replica computation\ucite{mezard1987spin} of the free energy. For systems with quenched disorder, this is a classical approach (the replicas or its equivalent formulation) to find the macroscopic behavior\ucite{barra2018phase,huang2017statistical,tubia2018,agliari2019free,hartnett2018replica}.

\subsection{Mean-field approach, the random-RBM}
\label{sec:rbm:rbm}
This MF approach to the macroscopic behavior of the RBM is based on statistical ensembles with iid elements of the weight matrix. Here, a random ensemble for the weight matrix is defined as follows. The weight matrix will be constructed using binary pattern: $w_{ia} = \frac{\xi_{ia}}{\sqrt{N_v}}$. Now, each pattern is selected to be
\begin{equation}
    \xi_{ia} = \left\{\begin{array}{lll}
         0 &  p_r \sim 1-p_i \\
         +1 &  p_r \sim p_i/2 \\
         -1 &  p_r \sim p_i/2 
    \end{array} \label{eq:dilution}\right.
\end{equation}

\noindent Using this definition, the degree of sparsity of the system is $p = \sum_i p_i/N_v$. The term random-RBM was coined by Tubiana et al.\ucite{tubiana2017emergence} but Agliari et al.\ucite{agliari2012multitasking,agliari2014multitasking} worked on a similar model although with a different theoretical approach. In particular, they computed the phase diagram in\ucite{barra2018phase}. We start by reproducing here the argument of Agliari that was developed for the RBM with a finite number of patterns before switching to the replica computation done in Tubiana's thesis\ucite{tubia2018}.

\begin{figure}
    \centering
    \includegraphics[scale=1.1]{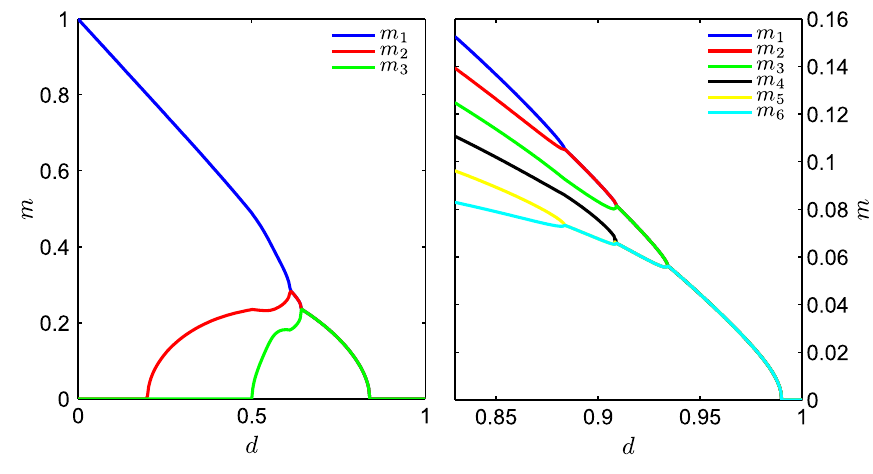}
    \caption{From\ucite{agliari2012multitasking}. Overlap with different patterns when varying the dilution factor $p$ (named $d$ on the figure) at low temperature. \textbf{Left:} a case with $3$ patterns where we can observe how at small dilution, only one pattern is fully retrieved while the second and third one appear for larger dilution. \textbf{Right:} a case with $6$ patterns where the figure is zoomed in the high dilution region where the branching phenomena is occurring and all the overlaps converge toward the same value.}
    \label{fig:agliari}
\end{figure}

\paragraph{Parallel retrieving ---} the usual definition of the Hopfield model (which we recall here is analogous to a Binary-Gauss RBM, see Section~\ref{sec:link_rbm}), consists in using extensive pattern $\xi_i^a = \pm 1/\sqrt{N_v}$ for all $i=1,\dots,N_v$, where $a=1,\dots,P$, $P$ being the number of patterns. The Hopfield model in the low storage regime, where the number of pattern is fixed, or scales logarithmically with the system size, is characterized by a low temperature regime made of configurations with an extensive overlap with \textit{one} of the patterns. This model can be recovered from a binary-binary RBM where the number of hidden nodes has the same scaling. Hence, having $N_h \sim \log(N_v)$, we can write exactly the partition function of the  binary-binary RBM in the limit of large system size 
\begin{align*}
  Z &= \sum_{\{\bm{s}\},\{\bm{\tau}\}} \exp\left( \beta \sum_{i,a} s_i w_{ia} \tau_a \right) \\
  &= \sum_{\{\bm{s}\}} \prod_a \cosh\left( \frac{\beta}{\sqrt{N_v}}\sum_i w_{ia} s_i \right) \approx \sum_{\{\bm{s}\}} \exp\left( \frac{\beta^2}{2 N_v} \sum_{i,j} \sum_a w_{ia} w_{ja} s_i s_j \right) \\       %= \sum_{\{\bm{s}\}} \exp\left( \frac{\beta^2}{2 N_v} \sum_a [\sum_{i} w_{ia} s_i]^2 \right) \\
  &= \sum_{\{\bm{s}\}} \exp\left( \frac{N_v \beta^2}{2} \sum_a m_a(\bm{s})^2 \right)
\end{align*}
recovering the Hopfield model with a square inverse temperature, and where we define the magnetization along the pattern $a$ as $m_a(\bm{s})$. In\ucite{agliari2012multitasking}, the authors considered a weight dilution as in eq. (\ref{eq:dilution}) applied to the above binary-binary RBM, or equivalently to a Hopfield model with a rescaled temperature. It is important to mention that it is a different procedure than diluting the network itself, see\ucite{wemmenhove2003finite} for more details on the other case. Having sparse patterns allows the network to retrieve more than one pattern at a time. In particular, global minima of the free energy can have an overlap with many patterns and locally stable states can be composed of a complex mixture of patterns.  We reproduce below on Figure~\ref{fig:agliari} the plot from the article\ucite{agliari2012multitasking} showing the overlap over three and six patterns in the (almost) zero temperature limit. We observe on the left panel that one pattern is fully retrieved when the dilution is low. Then, when increasing $p_i$, more and more patterns are retrieved together until the system enters a paramagnetic phase at high dilution.
%the retrieval phase is made of equilibrium states having an extensive overlap with one of the $P$ patterns. In\ucite{agliari2012multitasking}, a dilution parameter is included in the construction of the patterns: with a proba $p_i$, a component will be $\pm1$, and with $1-p_i$ it will be zero. It is important to mention that it is a different procedure than diluting the network itself, see\ucite{wemmenhove2003finite} for more details on the other case. Having sparse patterns allows the network to retrieve more than one pattern at a time. In particular, global minima of the free energy can have an overlap with many patterns and locally stable states can be composed of a complex mixture of patterns.  We reproduce below on Figure~\ref{fig:agliari} the plot from the article\ucite{agliari2012multitasking} showing the overlap over three and six patterns in the (almost) zero temperature limit. We observe on the left panel that one pattern is fully retrieved when the dilution is low. Then, when increasing $p_i$, more and more patterns are retrieved together until the system enters a paramagnetic phase at high dilution.

\paragraph{Replica approach of the random-RBM ---} We will now follow the approach of Tubiana et al.\ucite{tubia2018} and give more details on the derivation. This approach is based on a Bernoulli-RELU architecture giving the possibility to have continuous positive value for the hidden variables.

The characterization of the phase diagram is based on the determination of the free energy in thermodynamic limits. Given the weight ensemble (see eq. (\ref{eq:dilution})), the weight matrix is now made of independent and sparse elements. In this context, the replica analysis can be used to perform the quenched average. The replicated interaction term can be first easily computed and gives
\begin{equation*}
    \mathbb{E}_{\bm{w}} \left[ \exp\left(\beta \sum_{p} s_i^p w_{ia} \tau^p_a \right)\right] \approx \exp\left( \frac{p_i \beta^2}{2N} \sum_{pq} s_i^p s_i^q \tau_a^p \tau_a^q \right)
\end{equation*}

\noindent for the interaction term $(ia)$. The interaction between the visible and hidden nodes can be decoupled using the HS transformation

\begin{equation*}
  \exp\left( \frac{p_i^2 \beta^2}{2N} \sum_{pq} s_i^p s_i^q \tau_a^p \tau_a^q \right) \sim \int \prod_{pq} \frac{dq_{pq}d\bar{q}_{pq}}{2\pi} \exp\left( -N	\beta ( q_{pq} \bar{q}_{pq} - q_{pq} \frac{p_i}{p} s_i^p s_i^q - \frac{1}{2}\beta p \tau_a^p \tau_a^q )\right)
\end{equation*}
introducing the spin-glass order parameter over the replicas (we denoted by $p,q,\dots$, the replica index):
\begin{align*}
    \bar{q}_{pq} \sim& \mathbb{E}_{w}[\langle \tau_a^{q} \tau_a^{p} \rangle]\\
    q_{pq} \sim& \mathbb{E}_{w}[\langle (p_i/p) s_i^{q} s_i^{p} \rangle]
\end{align*}
where we note that the parameters over the visible nodes are weighted by the sparsity of the network. Using the replica symmetric ansatz, the quenched free energy can be computed and from it a set of order parameters emerges. An new order parameter is introduced in their derivation: the number $\tilde{L}$ of hidden nodes that have a macroscopic activation $\sim \mathcal{O}(m\sqrt{N})$, while the other ones remain silents (of order $1$). This parameter is reminiscent from the replica approach of the Hopfield model\ucite{AmGuSo1,AmGuSo2,AmGuSo3}. In this approach, the number of pattern that can be expressed is fixed, in order to investigate the stability of retrieving one or more patterns. The important difference here is that the sparsity $p$ imposes that the diluted weights can let many patterns to be expressed at the same time. Hence, the phase diagram will be characterized by, the value of the weight sparsity $p$ and $\tilde{L}$ the number of activated hidden nodes. The phase diagram is computed numerically, by scanning the possible value for the order parameters (see\ucite{tubia2018} for more details). It is found that when

\begin{itemize}
    \item $p=1$ and $\tilde{L}=1$: no sparsity and only one hidden node is activated. At low temperature, it gives back the behavior of the well-known Hopfield model having a recall phase of the patterns. An interesting additional result when using ReLu activations is that the capacity of the network can be increased by playing with the bias on the hidden nodes, at the cost of reducing the basin of attraction of the patterns.
    \item $p<1$ a ferromagnetic transition is found when imposing $\tilde{L}=1$, where one pattern is recalled at a time.
    \item $p<1$, when all the hidden nodes are all weakly activated, a SG phase is found.
    \item $p<1$ and $\tilde{L}$ is such that $1 \ll \tilde{L} \ll N_h$; a {\it compositional phase} is numerically identified. It is characterized by an intermediate number of hidden nodes strongly activated. %In practice, it is found solutions for $L>1$  by adjusting a parameter inhibiting or enhancing the activation of the hidden nodes.
\end{itemize}

\noindent In this analysis, it is demonstrated that in the possible \textit{equilibrium behaviors} of the random-RBM, an interesting phase mixing many patterns is present that characterizes in some way the efficient working regime of a learned  RBM. It is of course a simplified case where the patterns are $\{\pm1\}$ with a certain dilution factor. Now, the fact that there exists a family of weights where this phase exists is quite different from showing that the learning dynamics converges toward such a phase and how. In Tubiana's thesis, a stability analysis of the different phases is done showing that for a range of parameters of the RBM, the compositional phase is indeed the dominant one. Then, a certain number of numerical results are provided on the MNIST dataset which tends to confirm that the behavior of the learned RBM looks similar to a  "compositional phase". It would therefore be of great interest to characterize the learning curve theoretically in order to understand how this phase is reached. It is also interesting to mention a recent work investigating the role of the diluted weights\ucite{huang2018role} during the learning in a RBM with one hidden node. In this article, it is shown that the proportion of diluted weights tends to vanish during the learning procedure. This might be a signal that when the number of hidden features is very low, the RBM automatically adjusts itself in the ferromagnetic phase described above, learning a global pattern of the dataset.

\subsection{Mean-field approach using rank $K$ weight matrix}
\label{sec:svd:rbm}

The difficulty with the RBM is to be able to study the phase diagram of the model without discarding the fact that during the learning, the weights $w_{ia}$ become correlated between each others: starting from independently distributed $w_{ia}$, we can observe how the spectrum of the weight matrix is modified during the learning (see Figure \ref{fig:mnist_svd_modes} for instance). Classical approaches in statistical mechanics consider a set of independent weights, all identically distributed, before trying to compute the quenched free energy of the system by the replica trick\footnote{in few words, considering the quantity $Z^n$ for a given (integer) $n$, where $Z$ is the partition function, for small $n$, we can develop $Z^n \approx 1 + n\log(Z)$. The key point here is that it is generally possible to compute the quenched $Z^n$ and then making a small $n$ expansion.}. In the present case the hypothesis of independent weights cannot hold, as can be seen by looking at the spectrum of the weight matrix at the beginning of the learning and a few iterations later. The absorption of information by the machine prompts the development of strong correlations. This phenomena is illustrated in Section~\ref{sec:learning:svd_ica} on Figure~\ref{fig:mnist_svd_modes}. In order to understand how these eigenvalues affect the phase diagram of the system, it is reasonable to assume a particular statistical ensemble of the weight matrix of the form 
\begin{equation}
     w_{ia} = \sum_{\alpha=1}^K u_i^\alpha w_\alpha v_a^\alpha + r_{ia} \label{eq:form_w}
\end{equation}

\noindent where $K \ll N_v$, assuming a low-rank decomposition of the weight matrix plus some random noise $r_{ia}$, where $\bm{r}$ is a random  matrix with iid centered Gaussian elements with variance $\sigma$. With this decomposition we assume that the eigenvalues $w_\alpha$ correspond to some  intrinsic property of a learned dataset, while the matrix, $\bm{r}$, $\bm{u}$ and $\bm{v}$ can be treated as quenched disorder and averaged over. The set of vectors $\bm{u}^\alpha$ and $\bm{v}^\alpha$ correspond approximately to the left and right eigenvectors of the matrix $\bm{w}$. We can thus start to average $Z^n$ over all these variables. Starting with the average over the random matrix $\bm{r}$, it introduces the following interaction term $ \sum_{ia,p\neq q} s_i^{p} s_i^{q} \tau_a^{p} \tau_a^{q}$, where $p,q$ runs over the $n$ replicas. In this term, it is possible to decouple the interaction between the visible and the hidden nodes by introducing the overlap parameters 
\begin{align}
    Q_{pq} \sim& \mathbb{E}_{r,v,u}[\langle \tau_a^{q} \tau_a^{p} \rangle] \label{eq:def:ov1}\\
    \bar{Q}_{pq} \sim& \mathbb{E}_{r,v,u}[\langle s_i^{q} s_i^{p} \rangle] \label{eq:def:ov2}
\end{align}

\noindent Then, the form of the weight matrix, eq. (\ref{eq:form_w}) leads to the following change of variable% to decouple the interaction term between the $s$ and $\tau$ 
\begin{align*}
    s_\alpha = \frac{1}{\sqrt{L}} \sum_i s_i u_i^\alpha \\
    \tau_\alpha = \frac{1}{\sqrt{L}} \sum_a \tau_a v_a^\alpha
\end{align*}

\noindent where $L=\sqrt{N_v N_h}$. It corresponds to the projection of the visible and hidden variables over the matrix $\bm{u}$ and $\bm{v}$ coming from the SVD of $\bm{w}$. With this projection we will be able to define the order parameters of the system as the condensation of the visible and hidden nodes over the SVD modes of $\bm{w}$. Using again the Hubbard-Stratanovitch (HS) transformation in order to define the replicated magnetization
\begin{equation*}
    \exp\left( \sum_{ia} s_i^p w_{ia} \tau_a^p \right) = \exp\left( \sum_{\alpha} w_{\alpha} s_\alpha^p \tau_\alpha^p \right) \propto \int \prod_\alpha \frac{dm_\alpha^p d\bar{m}_\alpha^p}{2\pi} \exp\left(-L\sum_\alpha w_\alpha (m_\alpha^p \bar{m}_\alpha^p - m_\alpha^p s_\alpha^p - \bar{m}_\alpha^p \tau_\alpha^p) \right) \nonumber
\end{equation*}

\noindent we obtain two additional order parameters
\begin{align*}
    m_\alpha^p \sim& \mathbb{E}_{r,v,u}[\langle \tau_a^{p} \rangle]  \\
    \bar{m}_\alpha^p \sim& \mathbb{E}_{r,v,u}[\langle s_i^{p} \rangle]
\end{align*}

\noindent namely the condensation of the visible (resp. hidden) nodes along the SVD modes of $\bm{w}$. After some computation, the replicated free energy is obtained
\begin{align}
  \mathbb{E}_{u,v,r}[Z^n] &= \int \prod_{p,\alpha}\frac{dm_\alpha^p d\bar m_\alpha^p}{2\pi}\prod_{p\ne q}\frac{dQ_{pq}d\bar Q_{pq}}{2\pi} \nonumber\\[0.2cm]
&\times\exp\Bigl\{-L\Bigl(\sum_{p,\alpha}w_\alpha m_\alpha^p \bar{m}^p_\alpha+\frac{\sigma^2}{2}\sum_{p\ne q}Q_{pq}\bar Q_{pq}-\frac{1}{\sqrt{\kappa}}A[m,Q]
-\sqrt{\kappa}B[\bar m,\bar Q]\Bigr)\Bigr\} \label{eq:mf:FE}
\end{align}

\noindent where $\mathbb{E}$ indicates an average over the variables that are in subscripts and $\kappa=N_h/N_v$. The quantities $A$ and $B$ are given by
\begin{align}
A[m,Q] &\equiv \log\Bigl[\sum_{S^a\in\{-1,1\}}\mathbb{E}_u \Bigl(e^{\frac{\sqrt{\kappa}\sigma^2}{2}\sum_{p\ne q}Q_{pq}S^p S^q
+\kappa^{\frac{1}{4}}\sum_{p,\alpha}(w_\alpha m_\alpha^p -\eta_\alpha)u^\alpha S^p}\Bigr)\Bigr],\label{eq:Ap}\\[0.2cm]
B[\bar m,\bar Q] &\equiv \log\Bigl[\sum_{S^p\in\{-1,1\}}
\mathbb{E}_v \Bigl(e^{\frac{\sqrt{\kappa}\sigma^2}{2}\sum_{p\ne q}\bar Q_{pq}\tau^p \tau^q +\kappa^{-\frac{1}{4}}\sum_{p,\alpha}(w_\alpha \bar m_\alpha^p -\theta_\alpha)v^\alpha \tau^p}\Bigr)\Bigr].\label{eq:Bp}
\end{align}

\noindent In order to avoid more cumbersome computations we will skip the details, the interested reader being referred to\ucite{decelle2018thermodynamics}. The phase diagram of the model is based on the behaviour of the order parameters $Q$, $\bar{Q}$, $m$ and $\bar{m}$. After taking the saddle point of the free energy in the limit $L\to \infty$ keeping $\kappa$ fixed, using the replica symmetric ansatz and letting the number of replica go to zero, it is possible to distinguish different phases according to the values of the order parameters solutions to the saddle point equations. The order parameters of the systems in the replica symmetric phase are the condensation over the SVD modes (both for the visible and hidden nodes) $\hat{m}_\alpha$ and $m_\alpha$ and the overlaps $\hat{q}$ and $q$. The saddle point equations of the free energy leads to the following self-consistent equations for the order parameters:
\begin{align}
m_\alpha &= \kappa^{\frac{1}{4}}\mathbb{E}_{v,x}\Bigl[v^\alpha\tanh\bigl(\bar h(x,v)\bigr)\Bigr]\qquad\qquad
q = \mathbb{E}_{v,x}\Bigr[\tanh^2\bigl(\bar h(x,v)\bigr)\Bigr],\label{eq:mf1}\\[0.2cm]
\bar m_\alpha &= \kappa^{-\frac{1}{4}}\mathbb{E}_{u,x}\Bigl[u^\alpha\tanh\bigl(h(x,u)\bigr)\Bigr]
\qquad\qquad \bar q = \mathbb{E}_{u,x}\Bigl[\tanh^2\bigl(h(x,u)\bigr)\Bigr],\label{eq:mf2}
\end{align}
where
\begin{align*}
h(x,u) &=  \kappa^{\frac{1}{4}}\bigl(\sigma\sqrt{q}x+\sum_\gamma (w_\gamma m_\gamma-\eta_\gamma) u^\gamma\bigr),\\[0.2cm]
\bar h(x,v) &= \kappa^{-\frac{1}{4}}\bigl(\sigma\sqrt{\bar q}x+\sum_\gamma (w_\gamma\bar m_\gamma-\theta_\gamma) v^\gamma\bigr).
\end{align*}

\noindent A first look at the equations for the magnetization over the mode $\alpha$ tells us that they correspond to the usual mean-field equations of the Sherrington-Kirkpatrick model\ucite{kirkpatrick1978infinite} projected on the SVD decomposition of the weight matrix. The same is true for the overlap, with the difference that we have an overlap parameter for each layer. Analyzing these equations, we can distinguish three phases.
\begin{itemize}
    \item \textbf{A Paramagnetic phase:} it correponds to the case where $q=0$, $\hat{q}=0$, $m_\alpha = 0$ and $\hat{m}_\alpha = 0$. In the high temperature phase there exists only one minimum to the free energy.
    \item \textbf{A Ferromagnetic phase:} given by $q,\bar{q},m_\alpha,\bar{m}_\alpha \neq 0$. In this phase, the magnetization of the system is polarized toward one or many modes $\alpha$.
    \item \textbf{A Spin Glass phase:} where $q,\bar{q} \neq 0$, but $m_\alpha=\hat{m}_\alpha = 0$. In that phase, the system is trapped into one of the many minima of the free energy that are completely uncorrelated with the SVD modes of the weight matrix.
\end{itemize}

\begin{figure}
    \centering
    \includegraphics[scale=0.4]{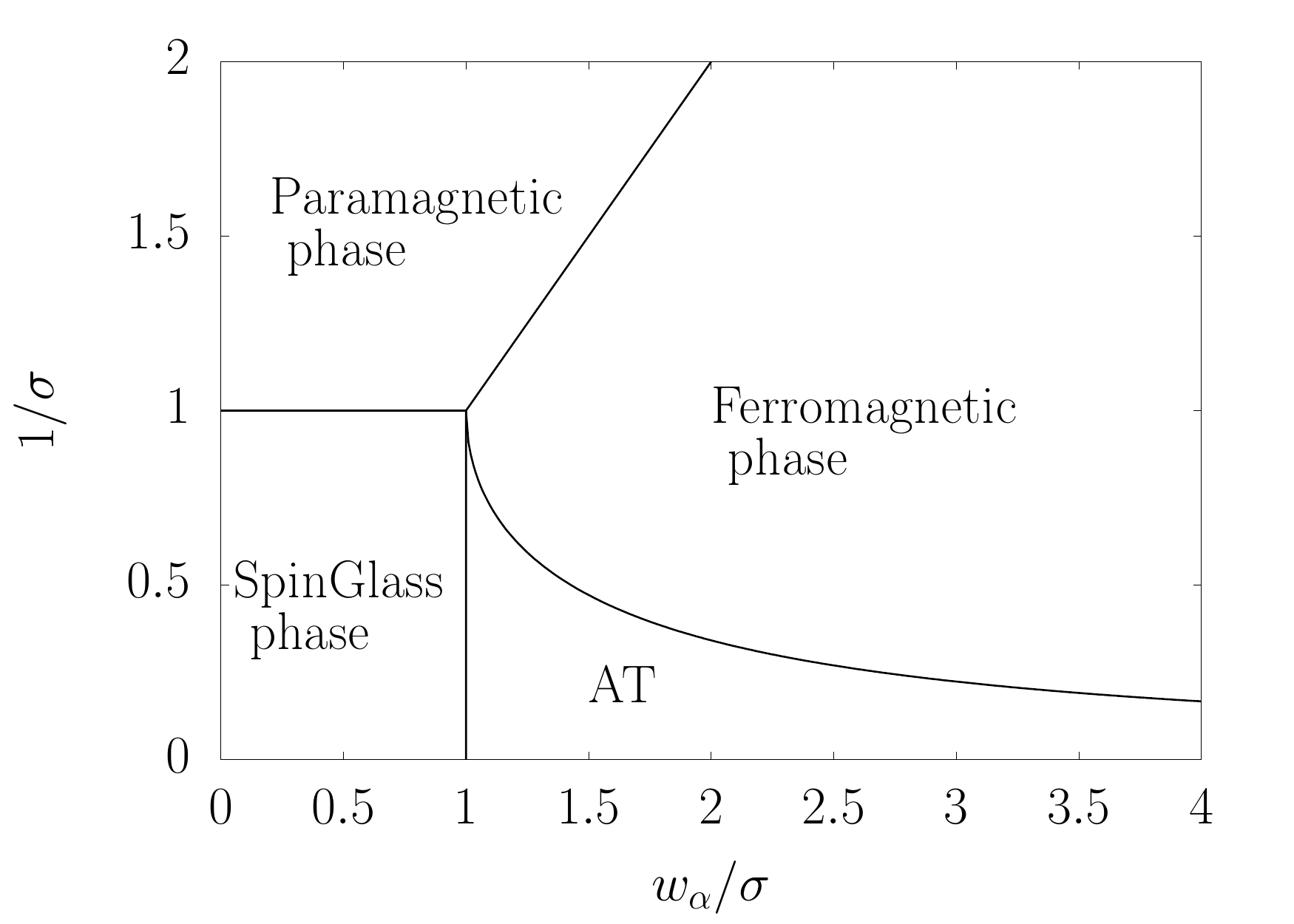}
    \includegraphics[scale=1.3]{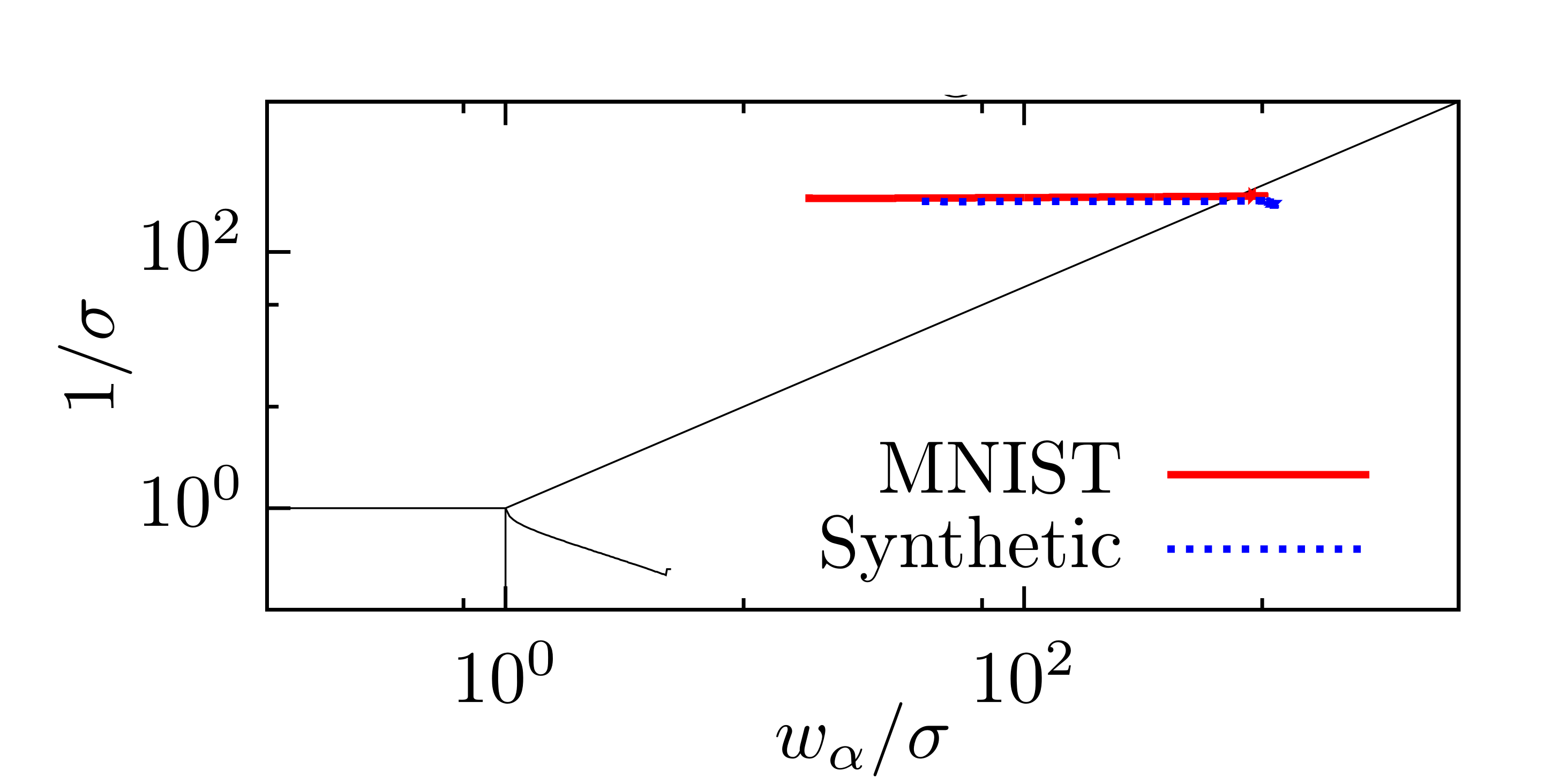}
    \caption{\textbf{Left:} the phase diagram of the model. The y-axis corresponds to the variance of the noise matrix, the x-axis to the value of the strongest mode of $\bm{w}$. We see that the ferromagnetic phase is characterized by having strong mode eigenvalues. In this phase, the system can behave either by recalling one eigenmode of $\bm{w}$ or by composing many modes together (compositional phase). For the sake of completeness, we indicate the AT region where the replica symmetric solution is unstable, but for practical purpose we are not interested in this phase. \textbf{Right:} An example of a learning trajectory on the MNIST dataset (in red) and on a synthetic dataset (in blue). It shows that starting from the paramagnetic phase, the learning dynamics brings the system toward the ferromagnetic phase by learning a few strong modes.}
    \label{fig:phasediag_svd}
\end{figure}

\noindent On left panel of Figure~\ref{fig:phasediag_svd} is shown the phase diagram as a function of $1/\sigma$  and of $w_{\rm max}/\sigma$, the ratio of strongest mode of $\bm{w}$ to the
variance $\sigma$ of the noise.

From the learning perspective, the  interesting phase is the ferromagnetic one. It seems also important that the learning avoid entering into the spin glass (SG) phase. The SG phase, apart from being uncorrelated with the SVD of $\bm{w}$, can affect very badly the MCMC that is used to compute the gradient. By inspecting the phase diagram on the left panel of Figure~\ref{fig:phasediag_svd}, we understand that at the beginning of the learning it is important to start with a weight matrix with a small variance $\sigma$ in order to avoid starting from the SG phase. Then, we expect during learning that one or many eigenvalues $w_\alpha$ will be expressed and that the trajectory will drift toward the ferromagnetic phase.

\paragraph{The nature of the ferromagnetic phase ---} it is instructive to look more in details at the ferromagnetic phase to understand the behavior of the RBM. We can distinguish two cases: in the first one only one eigenvalue $w_\alpha$ is learned ($w_\alpha > \sigma)$ and the other ones are close to zero; in the second scenario, many eigenvalues are expressed. In fact the first case is quite simple. Since only one mode has been learned the system will condense along this mode and it will be very similar to a ferromagnet. In the second scenario, we may have many $w_\alpha$ that have been learned, i.e. which are above noise threshold. The question then is whether the system will preferentially condense along one single mode taken out of the learned ones or whether it will be able to make compositions by condensing on several modes at the same time. In order to analyze this second scenario, it is important to recall that in order to derive the phase diagram one has to perform the quenched averaging over the matrices $\bm{u}$ and $\bm{v}$. The results will depend on the distribution that is used for the averaging. In\ucite{decelle2018thermodynamics}, it is shown that depending on the kurtosis of the distribution taken over $\bm{u}$ and $\bm{v}$, the system can behave in different ways. Denoting with $\gamma$ the relative kurtosis (w.r.t. the normal distribution) three different behaviors are identified:

\begin{itemize}
    \item $\gamma=0$, e.g. the Gaussian distribution. In that case, only the strongest mode is stable, and the weaker ones are unstable w.r.t. to the strongest one. Here, the system will condense along the strongest mode only.
    \item $\gamma<0$ e.g. the uniform or the Bernoulli distribution. Here the weaker modes can be metastable if they are not ``too far away'' from the strongest one. However the system will condense only toward one mode.
    \item $\gamma>0$ e.g. a sparse Bernoulli, or the Laplace distribution. In that case, the strongest mode is unstable w.r.t. weaker ones, leaving the possibility to have condensation over many modes at the same time. This corresponds to a dual compositional phase, by reference to the terminology introduced in~\ucite{tubiana2017emergence} which corresponds to combination of features instead of modes.
\end{itemize}
 
Hence depending on the form that will take the matrices $\bm{u}$ and $\bm{v}$ during the learning, different types of condensation may appear. This give us some insight on the way the statistical properties of the SVD of the weight matrix are reflected on the recall phase. In some cases the system might recall one macroscopic state, in another one an equilibrium state can be made of a mixture of modes. We illustrate on the right panel of Figure~\ref{fig:phasediag_svd} the learning trajectory on the phase diagram obtained both on artificial and MNIST data.

\section{Learning RBM}
\label{sec:learning_rbm}
Let us now discuss possible mechanisms at work during the learning of a RBM, which as we expect should have something to do with pattern formation mechanisms~\ucite{Amari}. We start by summarizing what is understood in exactly solvable models such as the Gaussian-Gaussian and Gaussian-Spherical RBMs. Then we will review a recent work\ucite{harsh2020place} showing how the learning dynamics on a simple dataset for the Bernoulli-Bernoulli RBM with one hidden node can be cast into a spatial diffusion equation. Then we will investigate numerically the behavior of the RBM on the MNIST dataset. In particular, how the learned features at short time can be interpreted using the SVD of the weight matrix and how, at later time, they seem to change completely. Then leaving aside the classical approach based on the Monte-Carlo computation of the gradient --- contrastive divergence\ucite{hinton2002training}, persistent contrastive divergence, \ucite{tieleman2008training}, parallel tempering\ucite{hukushima1996exchange,desjardins2010parallel,tubia2018}--- we will show how to use the MF self-consistent equations in order to compute the negative term to perform the learning. Finally, we will focus on the ensemble average equations for the learning, where we show how the MF theory developed in section (\ref{sec:svd:rbm}) can be integrated numerically and lead to the learning curve of the weight matrix $\bm{w}$.

\subsection{Learning dynamics for exactly solvable RBMs}

\begin{figure}
    \centering
    \includegraphics[scale=0.6]{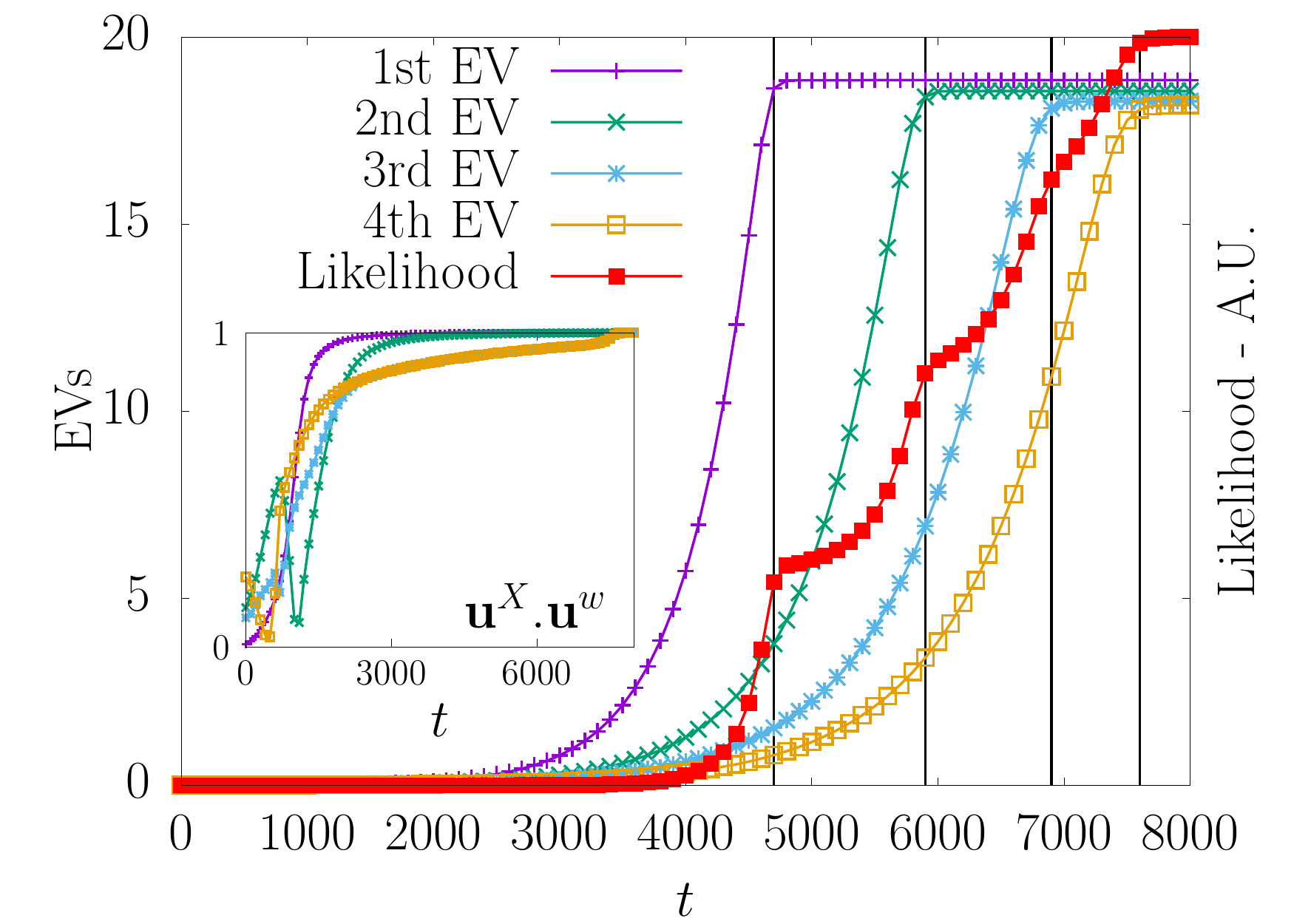}
    \caption{On this artificial dataset, we observe that eigenvalues that follows $\langle s_\alpha \rangle^2 > \sigma_v^2$ are learned and reach the threshold indicated by eq. (\ref{eq:w_max}). In the inset, the alignment of the first four principal directions of the matrix $\bm{u}^\alpha$ of the SVD of $\bm{w}$ and of the dataset. In red, we observe that the likelihood function is increasing each time that a new mode emerges.}
    \label{fig:dyn:gaussgauss}
\end{figure}

\paragraph{Gaussian-Gaussian RBM ---}
We have already seen, in Section~\ref{sec:gaussrbm} that the gradient of the Gaussian-Gaussian RBM can be computed exactly and how to characterize the growth of the eigenmodes of the weight matrix when freezing the rotation of the matrices $\bm{u}^\alpha$ and $\bm{v}^\alpha$. We put additional results here, operated on an artificial dataset\ucite{decelle2017spectral} containing $4$ well-separated Gaussian clusters. Recall that the modes of the SVD of the dataset that are higher than the intrinsic variance of the visible modes $\sigma_v^2$ will be expressed, and the vectors of rotation $\bm{u}^\alpha$ will aligned themselves with the principal directions of the dataset owing to eqs. (\ref{eq:sgd:walpha},\ref{eq:sgd_u},\ref{eq:sgd_v}). We can observe on Figure~\ref{fig:dyn:gaussgauss} the learning curve obtained for the first eigenmodes of the system coming out of the bulk. We can also see that the first eigenvectors $\bm{u}^\alpha$, associated to the expressed eigenvalues of $\bm{w}$, are aligning with the first principal directions of the SVD of the dataset. In parallel, we see that the likelihood --- that can be computed exactly here--- of the system increases stepwise after each new mode is learned.

\begin{figure}
    \centering
    \includegraphics[scale=1.0]{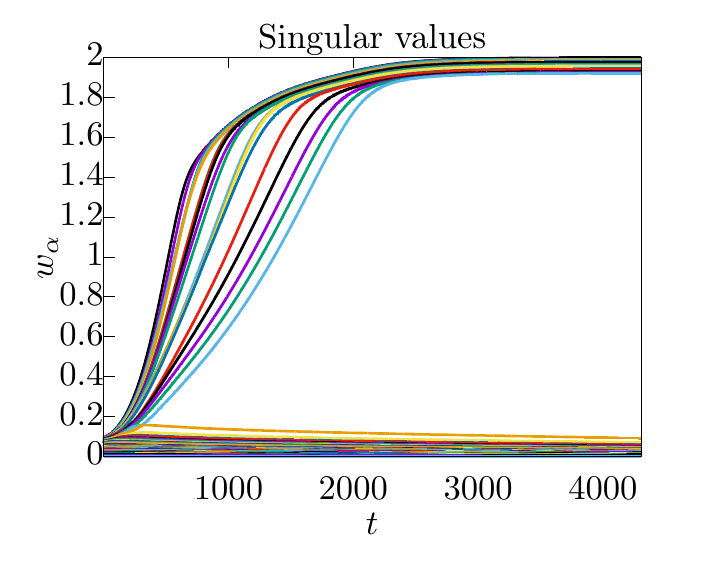}
    \includegraphics[scale=1.0]{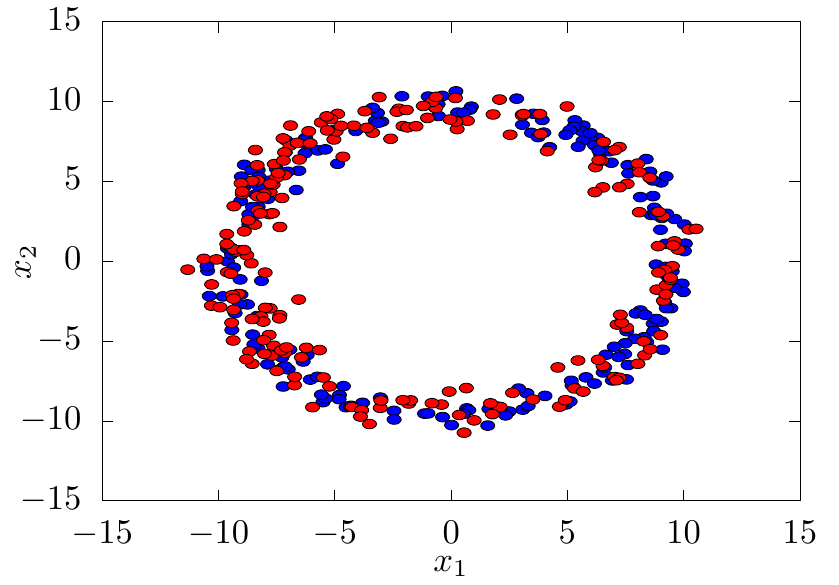}
    \caption{\textbf{Left:} the learning curves for the modes $w_\alpha$ using an RBM with $(N_v,N_h)=(100,100)$ learned on a synthetic dataset distributed in the neighborhood of a $20$d ellipsoid embedded into a $100$d space. Here the modes interact together: the weaker modes push the stronger ones higher, and they all accumulate at the top of the spectrum, as explained in Section~\ref{sec:sphrbm}. \textbf{Right:} a scatter plot projected on the first two SVD modes of the training (blue) and sampled data from the learned RBM (red) for a problem in dimension $N_v=50$ with two condensed modes. We can see that the learned matrix $\bm{u}$ captures relevant directions and that the RBM generates data perfectly similar to the one of the training set.}
    \label{fig:dyn:gausssph}
\end{figure}

\paragraph{Gaussian-Spherical RBM ---}
In the case of the Gaussian-Spherical case, it is again possible to obtain an exact analytical expression for the response function of the RBM $\langle s_\alpha \tau_\beta \rangle$~\ucite{decelle2020gaussian} for both the positive and negative terms, where the average is performed respectively over the dataset and the model distribution. The qualitative pictures is very similar to the previous one. As for the linear model, linear correlations between different modes vanish  and therefore the matrix $\bm{u}$ has to rotate until it is properly aligned with the principal directions of the dataset. At the same time singular values get either amplified or damped. In contrary to the linear case they do not evolve independently. Instead, as seen on the left panel of Figure~\ref{fig:dyn:gausssph} lower modes willing to condense exert some pressure on higher modes and accumulate at the top of the spectrum, hence pushing the whole spectrum upward. On the right panel of Figure~\ref{fig:dyn:gausssph}, to illustrate the result of mode condensation, we show a scatter plot containing data from the training dataset and data generated on the trained model when two modes condense.

\subsection{Pattern formation in the 1D Ising chain}

\begin{figure}
    \centering
    \includegraphics[scale=1.0]{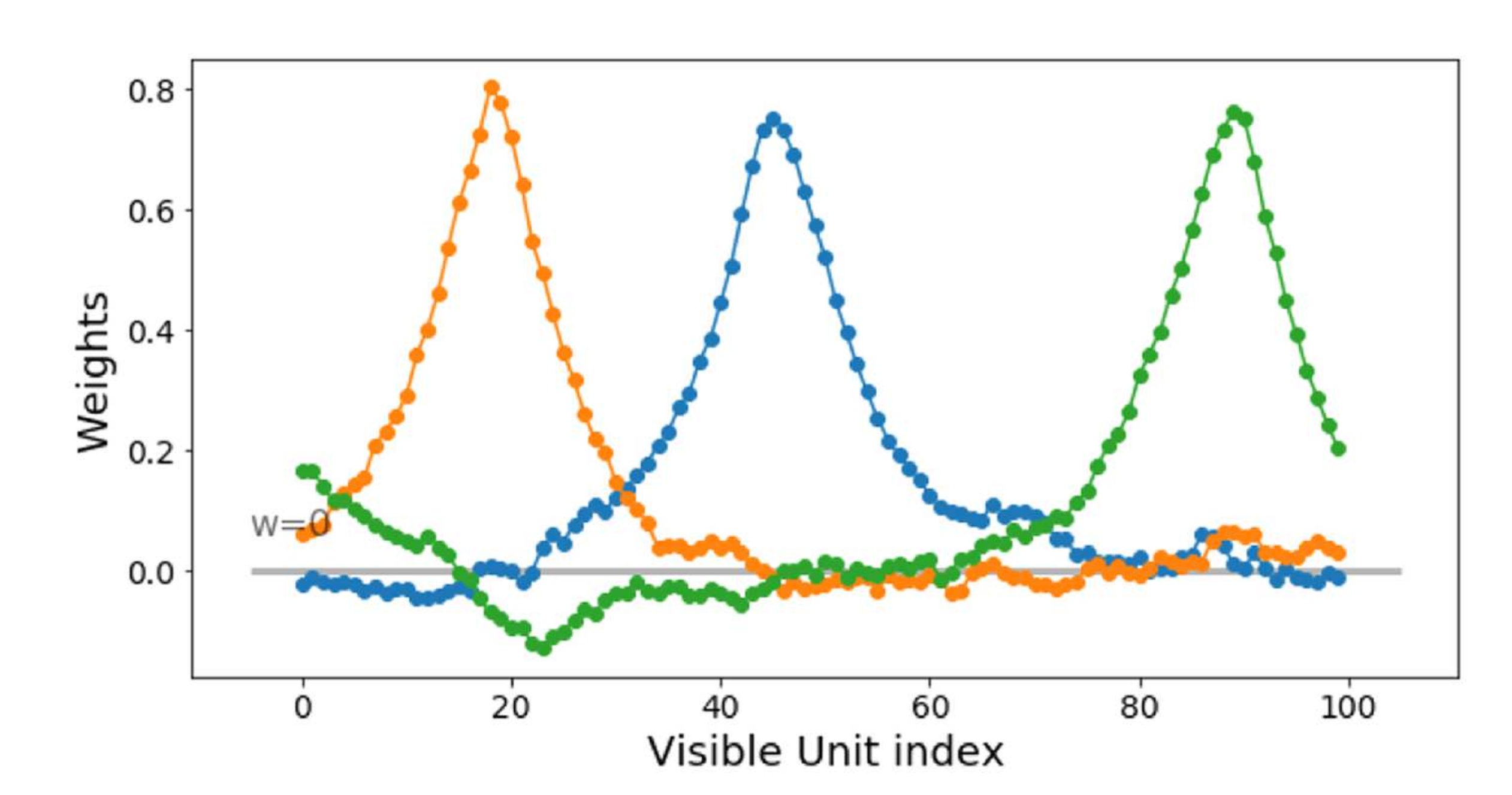}
    \includegraphics[scale=0.8]{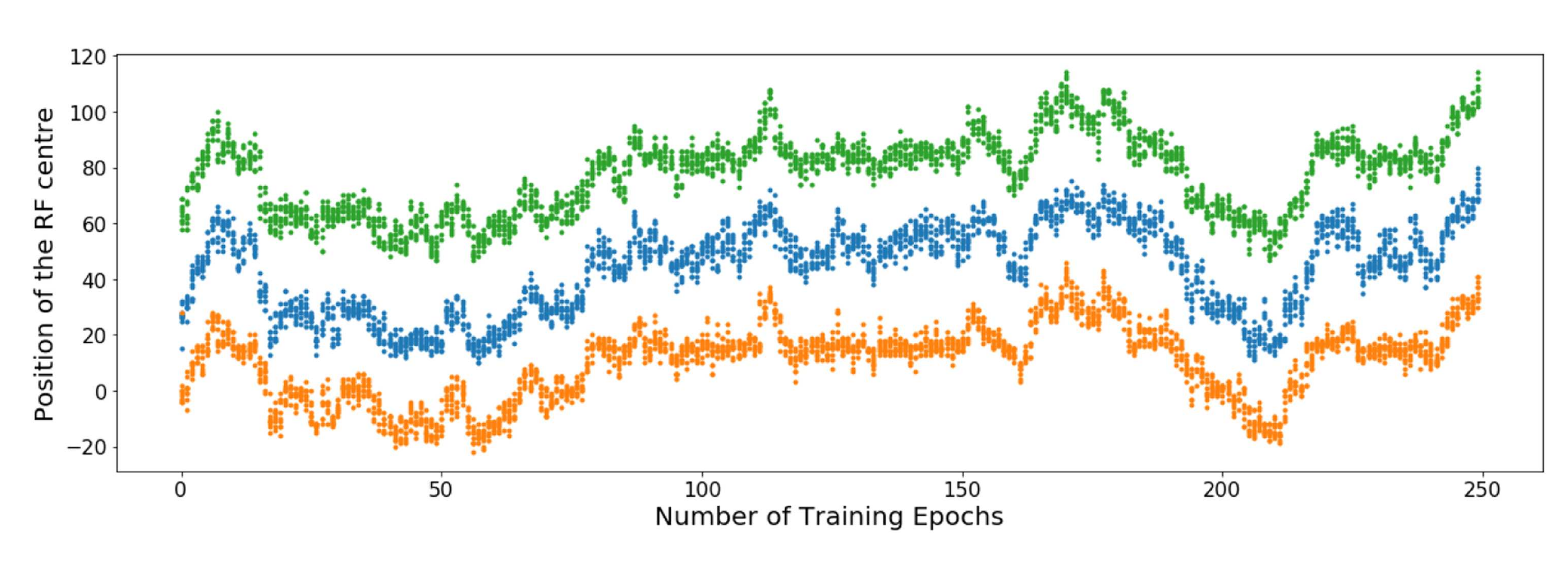}
    \caption{\textbf{Left:} figure from\ucite{harsh2020place}, the value of $w_i$ for each visible site of a RBM with $3$ hidden nodes trained on the dataset of the 1D homogeneous Ising model with periodic boundary condition. We see  three similarly peak shaped potentials with a decreasing magnitude of similar order for the three. Each peak intends to reproduce the correlation pattern around a central node, and therefore cannot reproduce the translational symmetry of the problem. \textbf{Right:} figure from\ucite{harsh2020place},  the position of the three peaks as a function of the number of training epochs. We observe that the  peaks diffuse while repelling each others. The diffusion aims at reproducing the correlation patterns of the translational symmetry, while the repelling interaction ensure that two peaks will not overlap.}
    \label{fig:1DIsing}
\end{figure}

In a recent work\ucite{harsh2020place}, the formation of features is studied analytically on a RBM with one hidden node. The training dataset is generated from a 1D Ising chain with a uniform coupling constant and periodic boundary conditions. The model used for generating the data has a translational symmetry which is exploited to solve the learning dynamics exactly. There is indeed available a closed form expression for the correlation function. Thanks to the translation invariance this depends only on the relative distance between the variables. Numerically, it is found that:
\begin{itemize} 
  \item the weights $\bm{w}$ function of the visible node index have a peak value for one of the visible node and decay with distance to this node. Since the position of the center breaks the translation symmetry it tends to diffuse over the system during the learning.
  
  \item Using more hidden nodes (but still few), it is observed that each feature is peaked at different places an repel each other to encode the correlation patterns of the data. Again, the position of the peaks diffuse with time even though some repulsive interaction seems to forbid them to cross. See Figure~\ref{fig:1DIsing} taken from\ucite{harsh2020place}  illustrating this phenomena.
\end{itemize}

\noindent Now in\ucite{harsh2020place}, the author compute the gradient of a system with one hidden node

\begin{equation*}
    \frac{\partial\log \mathcal{L}}{\partial w_i} = \left\langle s_i \tanh\left(\beta \sum_j s_j w_j \right) \right\rangle_{\rm data} - \tanh(w_i)
\end{equation*}

\noindent This expression can be developed up to the fourth order in $w$ (and $\beta$), giving in the case of the 1D Ising chain
\begin{equation*}
    \frac{\partial\log \mathcal{L}}{\partial w_i} \approx \beta ( w_{i+1} + w_{i-1})  - w_i \sum_k w_k^2 + w_i^3 + \mathcal{O}(w^4,\beta w^3)
\end{equation*}

\noindent \noindent It is easy to identify in the first two terms the 1D discrete spatial diffusion. This equation can be cast into a spatial diffusion equation with additional term in the continuous time limit (see\ucite{harsh2020place} for more details). From this small coupling expansion it is also possible to study the stationary solution in the one hidden node case and  show that it is consistent with experimental results: it describes a peaked function decreasing rapidly as the distance from the center increases. An approximated weak coupling equations can also be derived in the case of two hidden units. In that case, an effective coupling between the two features vectors $w_1$ and $w_2$ is present and responsible for a repulsive interaction between the two peaks.

This illustrates nicely how the features learned by the RBM tend to describe local correlations between variables. In addition, these features diffuse over the whole system during the learning to restore the translational symmetry without crossing thanks to a repulsive interactions between them. In the next section, we will focus on the learning behavior on the MNIST dataset and see that in that case, the learned features similarly describe local correlations.

\subsection{Pattern formation in MNIST: from SVD to ICA?}
\label{sec:learning:svd_ica}

\begin{figure}
    \centering
    \includegraphics[scale=8]{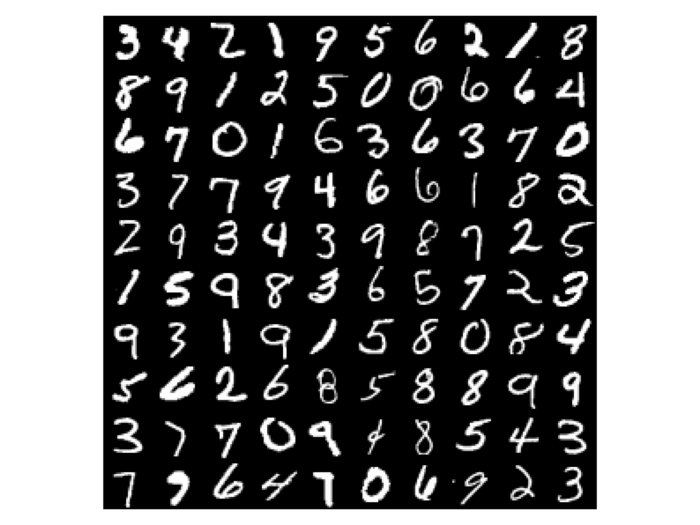}
    \caption{A subset of the MNIST dataset.}
    \label{fig:mnist}
\end{figure}

The pattern formation mechanism can be studied numerically on the MNIST dataset. MNIST\ucite{lecun1998gradient} is one of the most used real dataset in Machine Learning, it contains $60000$ images of black and white handwritten digits of $28 \times 28$ pixels, ranging from $0$ to $9$. The digits are about all the same size and are at the center of the image. They are illustrated on Figure (\ref{fig:mnist}).

To investigate how the patterns emerge from the learning process, we inspect the features during the learning on the Bernoulli-Bernoulli RBM. The first phase of the learning can be understood thanks to a standard linear stability analysis~\ucite{decelle2018thermodynamics,decelle2017spectral}. For this let us recall the learning behavior of the Gaussian-Gaussian RBM analyzed in Section~\ref{sec:gaussrbm}. In this simple case, the learning was triggered by the SVD of the dataset, and the growth of the modes $w_\alpha$ was controlled by how strong was the mode projected in the principal direction of the matrix $\bm{u}$.
%We will study numerically how the features develop in the case of MNIST
%, in particular knowing that the trigger of the learning is governed first by the strongest SVD modes of the dataset and that later on, more complex interactions will occur between the features, as demonstrated in the 1D Ising chain. Recall first the, in the Gaussian-Gaussin RBM (\ref{sec:gaussrbm}), the learning was 
%This phenomenology can also be understood as two different working mechanisms of the RBM. 
%In the mean-field description, the RBM can either condensate on a single mode. But, adjusting its parameters it can also recall a mixture of modes . Also, we can show that in the very weak coupling limit, the gradient tend to learn strongest modes of the SVD decomposition of the dataset, hence at the very beginning of the learning. 
Consider now the Bernoulli-Bernoulli RBM with $\{\pm1\}$ visible and hidden variables (to simplify), %\footnote{it differs from the $\{0,1\}$ case only by a shift in the local biases} 
and expand the log-likelihood gradient in the limit of small $\bm{w}$ (putting the local biases to zero):

\begin{align*}
    \frac{\partial \mathcal{L}}{\partial w_{ia}} &= \frac{1}{M} \sum_d s_i^{(d)} \tanh\left(\sum_j s_j^{(d)} w_{ja}\right) - \langle s_i \tau_a \rangle_{\mathcal{H}} \\
    &\approx \frac{1}{M} \sum_d s_i^{(d)} \sum_j s_j^{(d)} w_{ja} - w_{ia} \\
    &= \sum_j C_{ij} w_{ja} - w_{ia}
\end{align*}

\begin{figure}
    \centering
    \begin{minipage}{0.45\linewidth}
      \centering
      \includegraphics[scale=0.35]{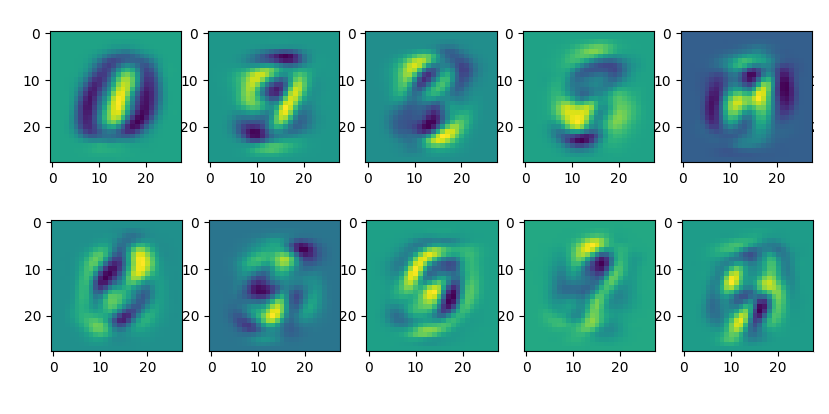}
      \includegraphics[scale=0.35]{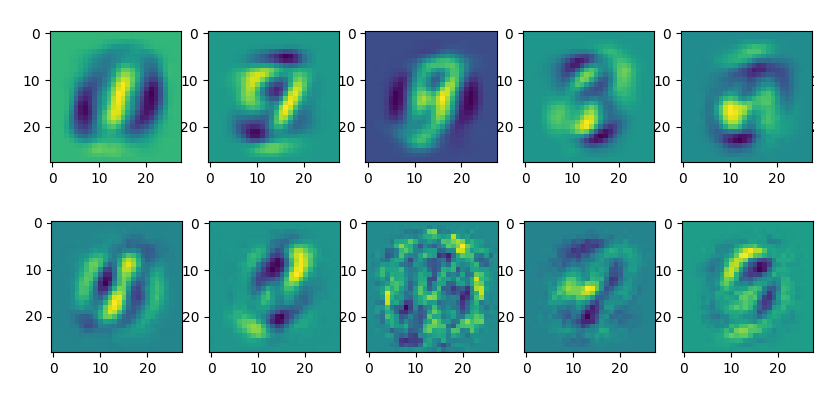}
    \end{minipage}
    \begin{minipage}{0.45\linewidth}
      \centering
      \includegraphics[scale=0.65]{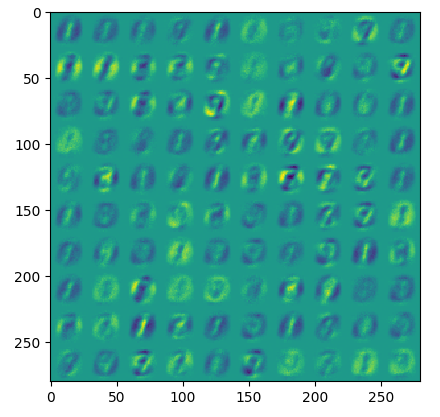}
    \end{minipage}
    \caption{\textbf{Left:} the first 10 modes of the MNIST dataset (top) and the RBM (bottom) at the beginning of the learning. The similarity between most of them is clearly visible. \textbf{Right:} $100$ random features of the RBM at the same moment of the learning. We can see that most features correspond to a mode of the dataset when comparing with the left-top panel.}
    \label{fig:modes_mnist_rbm}
\end{figure}

\noindent If we project these equations on the SVD modes of $\bm{w}$ as in Section~\ref{sec:gaussrbm}, we obtain the learning dynamics

\begin{equation*}
    \frac{dw_\alpha}{dt} = w_\alpha \left[ \langle \hat{s}_\alpha^2 \rangle - 1 \right],
\end{equation*}

\noindent identical at first order in $w_\alpha$ to (\ref{eq:sgd:walpha}) in the Gaussian-Gaussian case, when $\sigma_v=\sigma_h=1$. Hence, at the beginning of the learning, this RBM follows the same trajectory as the Gaussian-Gaussian one, where the mode $w_\alpha$ are amplified by the principal modes of the dataset. Similarly, it can be shown that the matrix $\bm{u}$ will start to align with the principal direction of the dataset. To see how the features evolve in the non-linear regime, we train an RBM with a very low learning rate and $500$ hidden nodes on MNIST. On Figure~\ref{fig:modes_mnist_rbm} we observe as expected from the linear stability analysis, that at the beginning of the learning the first modes of the weight matrix are almost identical to the one of the SVD of the dataset. We see in particular that the features themselves correspond to modes of the dataset, meaning that the RBM starts by learning global features.

\noindent Additionally, at this stage of the learning the MC samples obtained from the RBM are typically prototypes: each sample is almost identical (or have a large overlap) with a learned feature. In fact, during the training, if we monitor samples at each epoch (keeping a low learning rate), we can see that the samples have a high overlap with one mode at the beginning, then later on with combinations of modes. To be more precise, we can distinguish different stages of the learning by inspecting the features, the produced samples and the distance between the discretize features (taking the sign of each feature and computing the overlap).

\begin{figure}
    \centering
    \includegraphics[scale=0.45]{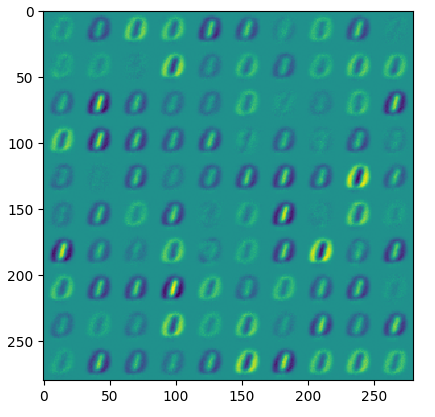}
    \includegraphics[scale=0.45]{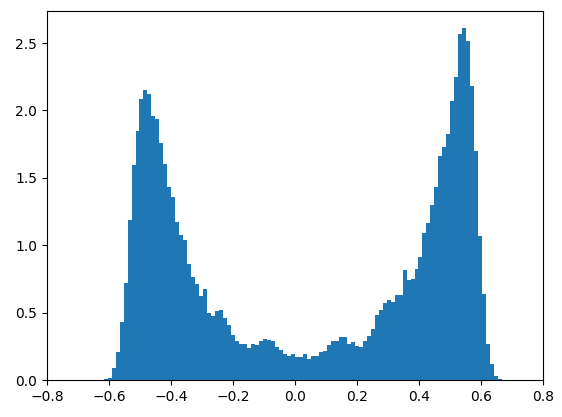}
    \includegraphics[scale=0.45]{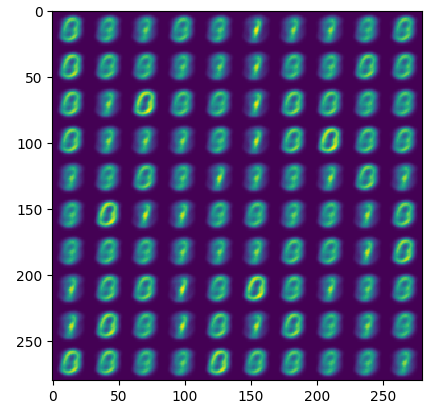}
    \includegraphics[scale=0.45]{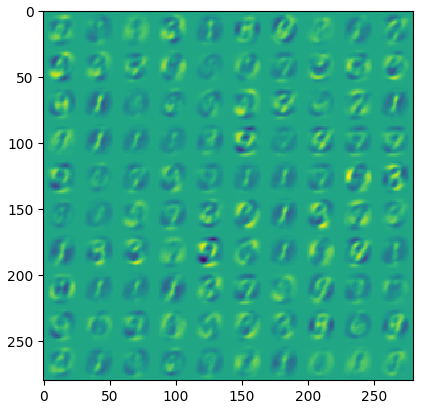}
    \includegraphics[scale=0.45]{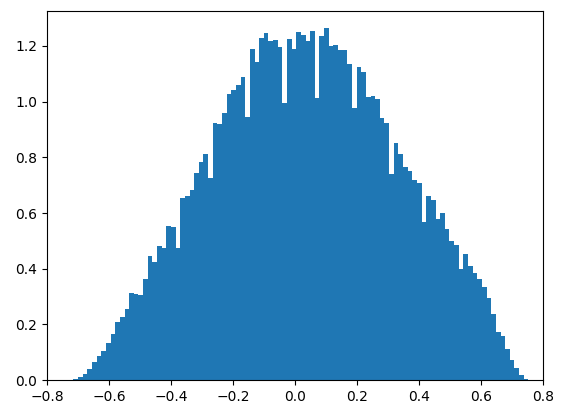}
    \includegraphics[scale=0.45]{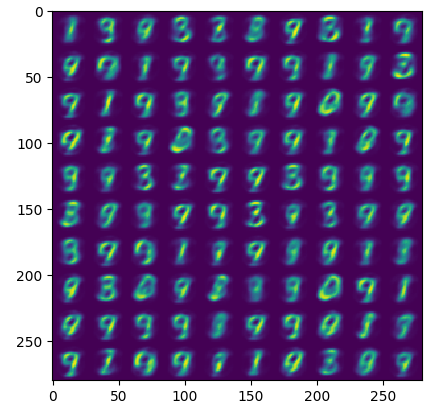}
    \includegraphics[scale=0.45]{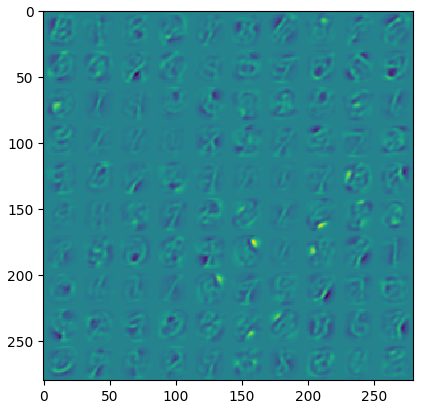}
    \includegraphics[scale=0.45]{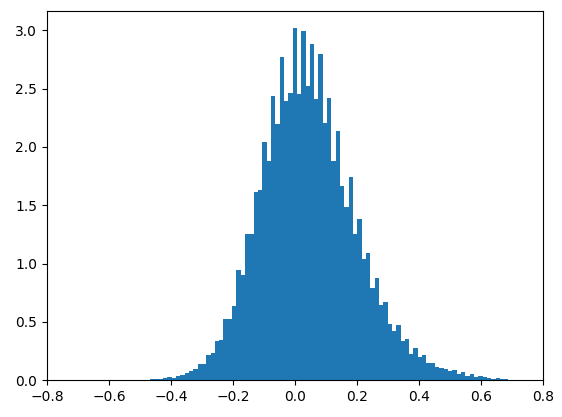}
    \includegraphics[scale=0.45]{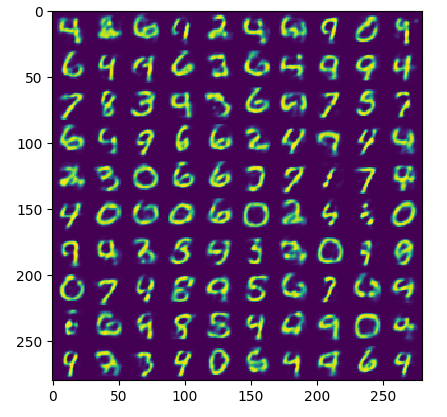}
    \caption{The column represents respectively  (i) the first hundred learned features, (ii) the histogram of distances between the binarized features: $W_{\pm 1} = \rm{sign}(W)$ and (iii) $100$ samples generated from the learned RBM. The first row corresponds to the beginning of the learning when only one feature is learned. Looking at the histogram, we see that most of the features have a high overlap. Also, the MC samples are all similar to the learned features. On the second row, the RBM has learned many features, and therefore the histogram is wider but still centered at zero. The MC sampling however is only capable of reproducing one of the learned features. On the last row the learning is much more advanced. The features tend to be very localized and the samples correspond now to digits.}
    \label{fig:learning_mnist}
\end{figure}

\noindent We illustrate these different stages on Figure~\ref{fig:learning_mnist}.

\begin{figure}
  \centering
  \begin{subfigure}{.32\linewidth}
    \includegraphics[width=\linewidth,trim=30 0 50 0,clip]{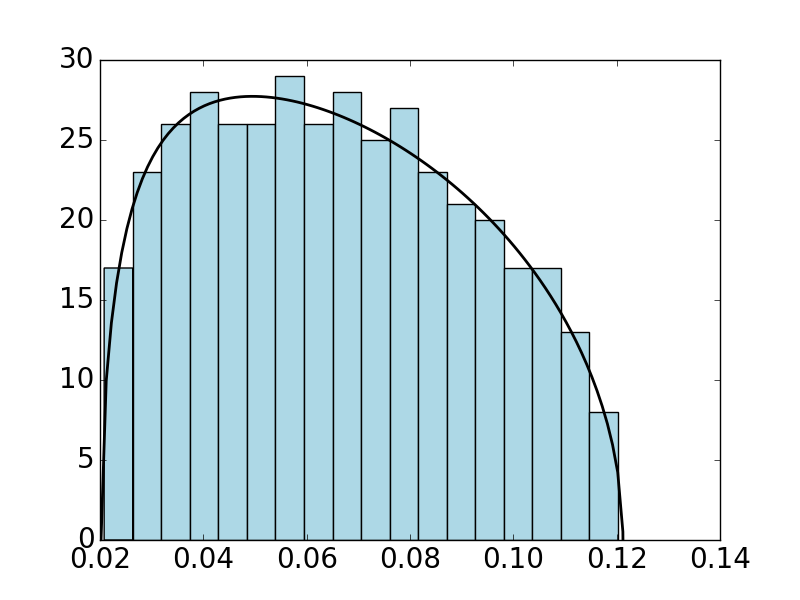}
    \caption{}
    \label{fig:mp_fit}
  \end{subfigure}
  \begin{subfigure}{.32\linewidth}
    \includegraphics[width=\linewidth,trim=30 0 50 0,clip]{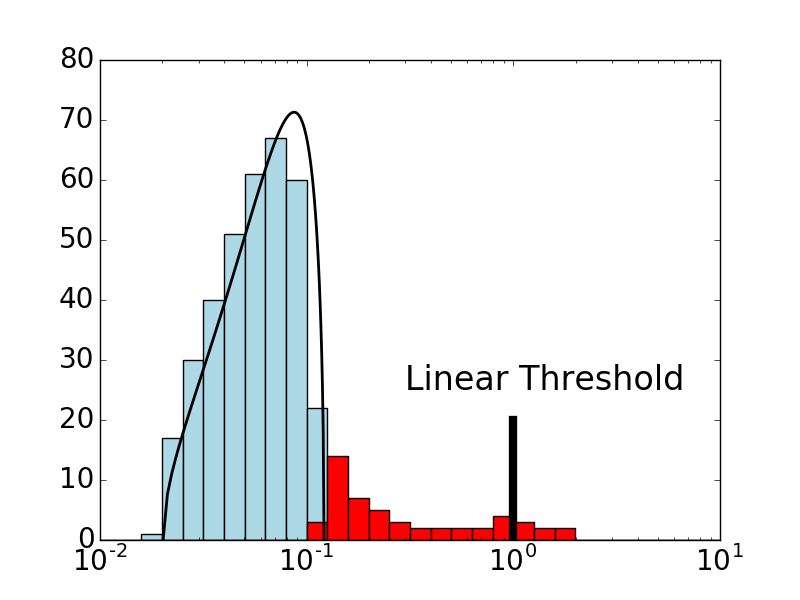}
    \caption{}
    \label{fig:sv1}
  \end{subfigure}
  \begin{subfigure}{.32\linewidth}
    \includegraphics[width=\linewidth,trim=30 0 50 0,clip]{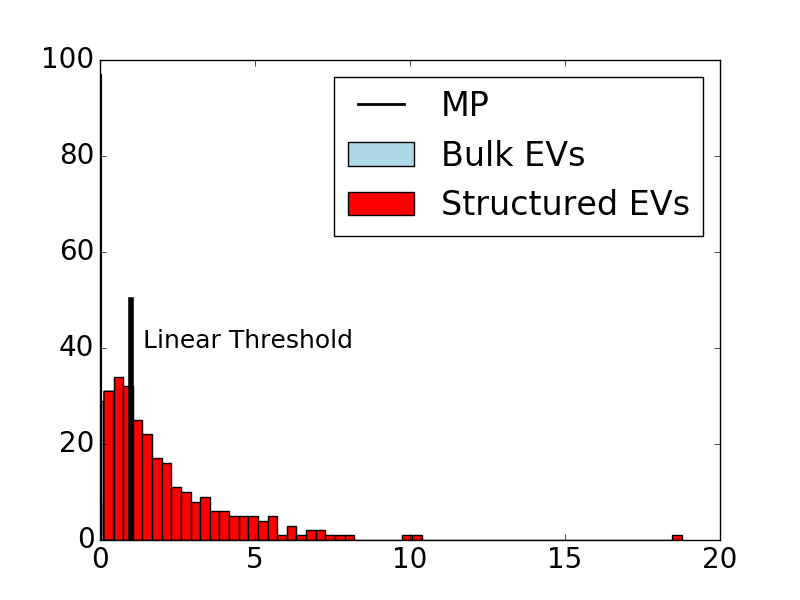}
    \caption{}
    \label{fig:sv2}
  \end{subfigure}\par\medskip
 \caption{\textbf{(a)} Singular values distribution of the initial random matrix compared to the Marchenko-Pastur law. \textbf{(b)} As the training proceeds we observe singular values passing above the threshold set by the Marchenko-Pastur law. \textbf{(c)} Distribution of the singular values after a long training: the Marchenko-Pastur distribution has disappeared and been replaced by a fat tailed distribution of eigenvalues mainly spread above threshold and a peak of below-threshold singular values near zero. The distribution of eigenvalues do not get close to any standard random matrix ensemble spectrum.}
 \label{fig:mnist_svd_modes}
\end{figure}

\noindent Finally at the end of the learning we recover localized features as in the study-case of the previous section. It has been noticed many times that these localized features are very similar to the ones given by an ICA. To which extent this aspect of learning is affected by the dataset that is considered is an open and interesting question. If we push further the learning, we observe that the RBM keeps learning more and more modes. It is not clear if the system enters in another phase (spin-glass or something else) or if it just overfits the dataset. To end up with these numerical experiments, let us look at the spectrum of $\bm{w}$, at the beginning, at an intermediate stage and at the end of the learning. On Figure~\ref{fig:mnist_svd_modes}, we see that starting from a Marchenko-Pastur law, coming from the spectrum of a Gaussian random matrix, quite quickly, many eigenvalues get out of the bulk as they are learned by the machine.

To summarize we have identified the following stages:
\begin{itemize}
    \item Stage 1: at initialization, the features are completely random and therefore the histogram of distances is Gaussian and centered at zero. The spectrum of $\bm{w}$ follows the Marchenko-Pastur distribution. The RBM starts from the paramagnetic phase.
    \item Stage 2: the RBM enters the ferromagnetic phase, the first strongest mode of the SVD is learned by all features, giving a high positive or negative overlap in the inter-features distances while the generated samples have a high overlap with the learned features.
    \item Stage 3: where many modes have emerged, but the learned features remain global and close to the modes of the dataset. The histogram of distances becomes much broader but the generated sample correspond basically to the learned features with few variety. The RBM is in a pure Mattis phase analogous to the recall phase of the Hopfield model.
    \item Stage 4: finally, after a much longer period, we observed that the learned features are much alike an ICA decomposition while the distances between features is still centered in zero but with a much smaller variance. Finally the generated samples look very similar to the provided dataset. The RBM is in a compositional phase, both regarding the features and the modes (the dual one). 
    \item Stage 5: empirically, we observe that the learning of the modes of $\bm{w}$ never stops. Hence, a macroscopic number of modes is expressed and it is not clear anymore what would be the behavior of the machine in this regime, whether this corresponds to a standard spin-glass phase\ucite{hartnett2018replica} or another unknown disordered phase.
\end{itemize}

In future works, it could be interesting to understand the mechanism leading to the localization of the features, in particular whether this is related to some specific tail distribution of the weight matrix spectrum. An aspect of RBMs completely absent from the previous description of the learning process is the behaviour of the biases associated to hidden nodes. These are very important since they determine the threshold above which the features are activated and their learning dynamics is quite intertwined with the modes dynamics. This aspect of the learning could be worth studying especially to improve present learning algorithms. 

\subsection{Learning RBM using TAP equations}
\label{sec:tap_learning}

The difficulty of learning an RBM comes as already said from the negative term which requires to compute the thermal average of correlations between a visible and hidden nodes. In particular, when the machine starts to learn many modes, it becomes more and more difficult to estimate this term correctly using Monte-Carlo methods due to the eventually large relaxation time. In addition, to get a precise measurement, it is necessary to get many statistically independent samples in order to reduce the statistical error. 

In this section we will derive the mean-field self-consistent equations that can be used to approximate the negative term by using a high-temperature expansion of the Boltzmann measure. We illustrate the method showing the result of Gabrié et. al\ucite{gabrie2015training} where a RBM has been trained by using the TAP equations. An interesting derivation using a variational approach in the case of the Gaussian-Bernoulli case has also been done in\ucite{takahashi2016mean}.

\paragraph{High-Temperature (Plefka) expansion ---}
We review here a famous approach using a high-temperature expansion of the system in order to compute the mean-field magnetization.
This method is both very simple to implement and also provides a way to approximate the free energy of the system in the weak couplings regime. Recent successful approaches\ucite{gabrie2015training,tramel2018deterministic} showed how it is possible to train a RBM using these mean-field equations. For this subsection, we will use $\{\pm1\}$ binary variables for simplicity. 

For the Ising model, it is well-known that the (naïve) mean-field (nMF) approximation can be written as a set of self-consistent equations on the magnetizations, and the associated approximation of the free energy can be computed as a function of these magnetizations: \

\begin{align}
  m_i &= \tanh\left( \sum_j J_{ij} m_j + h_i \right) \text{, } \forall i \nonumber\\
  F[\bm{m}] &= \sum_i \left[ \left( \frac{1-m_i}{2} \right) \log \left( \frac{1-m_i}{2} \right) + \left( \frac{1+m_i}{2} \right) \log \left( \frac{1+m_i}{2} \right)\right] \nonumber\\
  & - \sum_{i<j} J_{ij} m_i m_j - \sum_i m_i h_i \nonumber
\end{align}

\noindent These equation can be translated directly to the case of the RBM, with the only need to specify clearly which variables are the visible and hidden ones. One gets the following:

\begin{align}
  m_i &= \tanh\left( \sum_a w_{ia} m_a + \theta_i \right) \nonumber\\
  m_a &= \tanh\left( \sum_i w_{ia} m_i + \eta_a \right) \nonumber\\
  F[\bm{m}] &= \sum_i \left[ \left( \frac{1-m_i}{2} \right) \log \left( \frac{1-m_i}{2} \right) + \left( \frac{1+m_i}{2} \right) \log \left( \frac{1+m_i}{2} \right)\right] \nonumber\\
  &+ \sum_a \left[ \left( \frac{1-m_a}{2} \right) \log \left( \frac{1-m_a}{2} \right) + \left( \frac{1+m_a}{2} \right) \log \left( \frac{1+m_a}{2} \right)\right] \nonumber\\
  & - \sum_{i,a} w_{ia} m_i m_a - \sum_i m_i \eta_i - \sum_a m_a \theta_a \nonumber
\end{align}

\noindent Here, we remind the reader that we use the indices $i,j,k$ for the visible nodes, and $a,b,c$ for the hidden ones. We recognize on the first two lines of the free energy the entropy terms  $S(m_i)$ and $S(m_a)$ of the model for respectively the visible and the hidden nodes. Note first that the nMF approximation corresponds to a first order development in $\beta$ (or in small $\bm{w}$), but it can be generalized to higher orders, recovering the so-called TAP\ucite{thouless1977solution} equations at the second order. Second, we can generalize this scheme to any order using the Pfleka expansion\ucite{plefka1982convergence,georges1991expand}. Let us demonstrate first how to obtain the first and second order approximation in the case of $\pm1$ variables. To simplify the computation, we center all the terms around their mean value and make the computation for a case without local bias 

\begin{align*}
    \mathcal{H} &= -\sum_{i,a} s_i w_{ia} \tau_a \\
    &= -\sum_{ia}(s_i - m_i) w_{ia} (\tau_a - m_a) - \sum_i (s_i-m_i) \sum_a w_{ia} m_a \\ 
    & - \sum_a (\tau_a-m_a) \sum_i w_{ia} m_i - \sum_{ia} m_i w_{ia} m_a 
\end{align*}

\noindent Using this expression, we can follow\ucite{georges1991expand} and compute  the magnetization in the infinite temperature limit of the following free energy

\begin{equation*}
    -\beta A = \log\left[ \sum_{\{s,\tau\}} \exp(-\beta \mathcal{H} + \sum_i \lambda_i(\beta)(s_i - m_i) + \sum_a \lambda_a(\beta)(\tau_a - m_a) \right]
\end{equation*}

\noindent The relation between the magnetization and the Lagrange multipliers $\lambda$ are obtained by imposing  $m_i = \langle s_i \rangle_{\beta=0} = \lambda_i(0)$ and similar constraints for the hidden nodes. Then, we expand the free energy in a high temperature series

\begin{equation*}
    -\beta A = -\beta A\bigg\rvert_{\beta=0} - \beta \frac{\partial\beta A}{\partial \beta}\bigg\rvert_{\beta=0} - \frac{\beta^2}{2}\frac{\partial^2\beta A}{\partial \beta^2}\bigg\rvert_{\beta=0} + \dots
\end{equation*}

\noindent With our Hamiltonian, we can compute the first and second order easily

\begin{align*}
    %-\beta A \bigg\rvert_{\beta=0} &= -\sum_i \left( \frac{1-m_i}{2} \right) \log \left( \frac{1-m_i}{2} \right) + \left( \frac{1+m_i}{2} \right) \log \left( \frac{1+m_i}{2} \right) \\
    -\frac{\partial \beta A}{\partial \beta}\bigg\rvert_{\beta=0} &= \langle \mathcal{H} \rangle = \sum_{ia} m_i w_{ia} m_a \\
    -\frac{\partial^2 \beta A}{\partial \beta^2}\bigg\rvert_{\beta=0}&= \frac{1}{2} \sum_{ia} w_{ia}^2 (1 - m_i^2) (1-m_a^2)
\end{align*}

\noindent where we used the following identities for the second order computation

\begin{align*}
    \frac{\partial^2 \beta A}{\partial \beta \partial m_i}\bigg\rvert_{\beta=0} &= -\sum_a w_{ia} m_a \\
    \frac{\partial^2 \beta A}{\partial \beta \partial m_a}\bigg\rvert_{\beta=0} &= -\sum_i w_{ia} m_i
\end{align*}

\noindent As show in\ucite{gabrie2015training}, the expansion can be easily extended to the third order without a big computational cost due to the particular topology of the RBM. Deriving the free energy obtained at this order w.r.t. the magnetization, we obtain the self-consistent set of equations defining the TAP equations for the RBM

\begin{align}
    m_i = \tanh\left(\sum_a w_{ia} m_a - \sum_a w_{ia}^2 m_i (1- m_a^2) \right) \label{eq:tap_1}\\
    m_a = \tanh\left(\sum_i w_{ia} m_a - \sum_i w_{ia}^2 m_a (1- m_i^2) \right) \label{eq:tap_2}
\end{align}

\noindent Hence, a solution of the TAP equations should satisfied eqs. (\ref{eq:tap_1}) and (\ref{eq:tap_2}) and give us at the same time the approximated free energy associated to this solution:
\begin{equation}
    F[\bm{m}] = \sum_i S(m_i) + \sum_a S(m_a) - \sum_{ia} w_{ia} m_i m_a + \sum_{ia} \frac{w_{ia}^2}{2}\left(1 - m_i \right)^2\left(1 - m_a \right)^2
    \label{eq:tap:FE}
\end{equation}
We can now use these mean-field equations to learn the RBM. First, we need to take into account the fact that many solutions to eqs. (\ref{eq:tap_1}) and (\ref{eq:tap_2}) exist, each one with a given value  of the free energy. Hence, the partition function can be approximated by 

\begin{equation*}
    Z = \sum_{\gamma} e^{-F(m_i^{(\gamma)},m_a^{(\gamma)})}
\end{equation*}

\noindent where the sum runs over all the possible solution to the mean-field equations (\ref{eq:tap_1})-(\ref{eq:tap_2}), weighted by the free energy given in (\ref{eq:tap:FE}). Using this approximation in the computation of the likelihood we obtain the following gradient 

\begin{equation*}
    \frac{\partial \mathcal{L}}{\partial w_{ia}} = \langle s_i \tau_a \rangle_{\rm data} - \langle m_i m_a + w_{ia}^2 (1-m_i^2)(1-m_a^2) \rangle_{\rm MF}
\end{equation*}

\noindent where 

\begin{equation}
    \langle O \rangle_{MF} = \frac{\sum_\gamma O_\gamma e^{-F_\gamma}}{\sum_{\gamma'} e^{-F_{\gamma'}}}
\end{equation}

\noindent correspond to the model average over all the solutions of the mean-field equations. We can see here a notable difference with the approach developed in\ucite{gabrie2015training}. In their work, Gabrié et al runs the sums over all obtained fixed point from the mean-field equations divided by the number of fixed points only. The risk is that if the mean-field equations converge toward a fixed point that is suboptimal (have a high free energy) or even spurious (being a a maximum of the free energy) the estimation of the negative term will be polluted by such fixed points. More details on the Plefka expansion on bipartite Ising model can be found here\ucite{maillard2019high}. As a final remark, let us insist on the fact that, even if the convergence of the TAP equations is not guaranteed, problems of convergence are practically not met in the ferromagnetic phase. On the contrary, such problems occur quite often in the spin glass phase which we wish to avoid in the context of learning the RBM.

\paragraph{Experiment with TAP learning}

\begin{figure}
    \centering
    $\vcenter{\hbox{\includegraphics[scale=0.85]{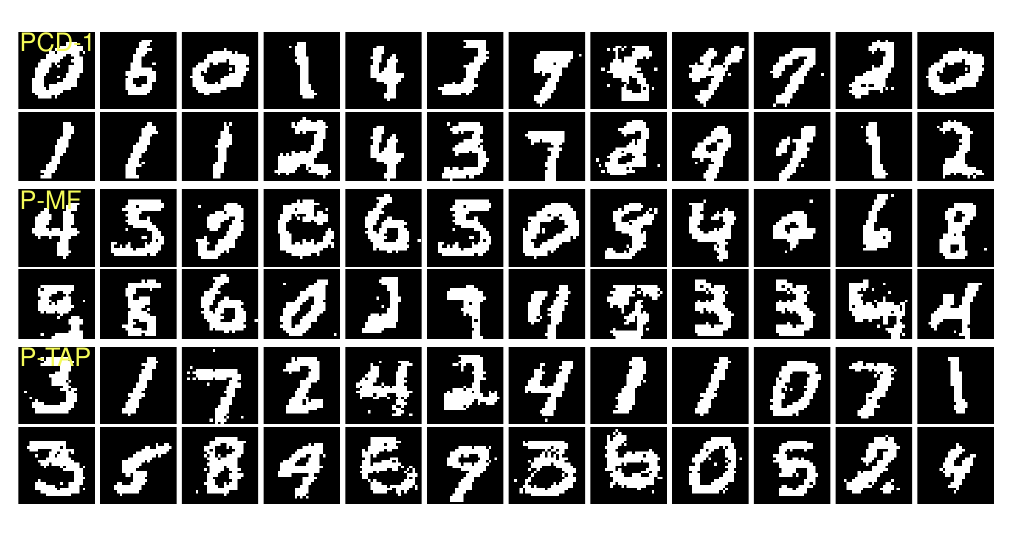}}}$
    $\vcenter{\hbox{\includegraphics[scale=0.55]{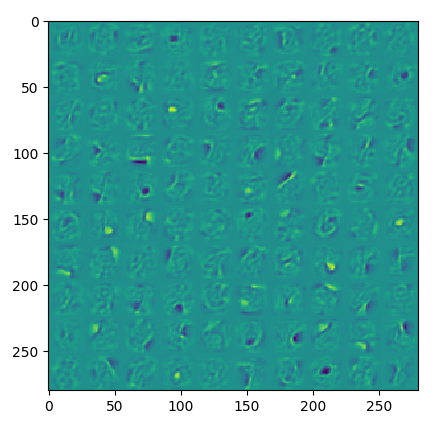}}}$
    \caption{\textbf{Top:} ---figure taken from\ucite{gabrie2015training}--- Samples taken from the permanent chain at the end of the training of the RBM. The first two lines correspond to samples generated using PCD. The second two lines to samples obtained using the P-nMF approximation and the last two, using P-TAP. \textbf{Bottom:} A $100$ features obtained after the training, we can see that they are qualitatively very similar to the ones obtain when training the RBM with P-TAP.}
    \label{fig:gabrie:TAP}
\end{figure}

We show here some results obtained on MNIST using the same parameters as above but with the mean-field approximation taken from\ucite{gabrie2015training}. Here, the comparison is done using the persistent chain algorithm, where a set of MC chains is maintained all along the learning whenever using CD, nMF or the TAP approximation (in the case of nMF or TAP, the chain is updated using the corresponding self-consistent equations), see Figure~\ref{fig:gabrie:TAP}. %In this particular setting however, it is important to shuffle the chain once and a while in case the obtain fixed points remain locally stable. To avoid this problem, a subpart of the chain is reinitialized at each epoch, either on random initial condition either using random datapoints. 

First we see that the samples generated by all the three methods are qualitatively similar. Second, the features learned are also qualitatively similar to the PCD case. Therefore, on the MNIST dataset the two machines are hardly distinguished by just looking at the generated samples and learned features indicating that the MF/TAP approximation is working very well. It is also important to point out here that, the advantage of the mean-field approximation in that case does not rely on any speed up with regard to the learning procedure. But, more importantly, it provides complementary tools such as the fixed points as local maxima of the free energy and their associated free energy. For instance, in\ucite{TramelIEEE} a RBM is used as a prior distribution in the context of compressed sensing where the mean-field equations are used to infer equilibrium values of the variables. In\ucite{fissore2019robust}, the RBM is used to reconstruct images from partial observations, again using the mean-field formulation to infer the states of the missing information.

\subsection{Mean-Field learning: ensemble average}

The mean-field equations derived in section (\ref{sec:svd:rbm}) for the RBM, where the weight matrix is constructed as a low rank decomposition, can be integrated numerically in order to learn the parameters of the RBM. By contrast to the TAP equations described in section (\ref{sec:tap_learning}) which are solved on single instances, they correspond to the ensemble average (over the parameters $\bm{u}$, $\bm{v}$ and the noise), i.e. are meant to represent an average case of learning. 

In the approach developed in\ucite{decelle2018thermodynamics}, using the statistical ensemble defined in Section~\ref{sec:svd:rbm} it is possible to have a mean-field estimate of the response functions involved in the gradient of the log-likelihood. For the response term on the data we get 
\begin{equation*}
    \langle s_\alpha \tau_\beta \rangle_{\rm data} = \langle s_\alpha (s_\beta w_\beta - \theta_\beta)(1-q_\beta[\bm{s}])\rangle_{\rm data} 
\end{equation*}

\noindent where the parameter $q_\beta[\bm{s}]$ is a variant attached to mode $\beta$ of the spin-glass parameter taken as a function of $\bm{\bar m}$ in equation (\ref{eq:mf1}),  when the visible nodes are pinned to the dataset (see~\ucite{decelle2018thermodynamics} for details). The negative term is more complicated to compute. It depends on the fixed points obtained through equations (\ref{eq:mf1}) and (\ref{eq:mf2}) for a given set of model parameters. Once the fixed point are obtained, the response terms of the RBM can be written 

\begin{align*}
    \langle s_\alpha \tau_\beta \rangle_{\rm {\mathcal H}}& = \frac{1}{Z_{\rm MF}} \sum_{\gamma} e^{-L f(m^\gamma,\bar{m}^\gamma,q^\gamma,\bar{q}^\gamma)} m_\alpha^\gamma \bar{m}_\beta^\gamma = \langle m_\alpha^\gamma \bar{m}_\beta^\gamma \rangle_{\rm MF} \\
    Z_{\rm MF} &= \sum_{\gamma} e^{-L f(m^\gamma,\bar{m}^\gamma,q^\gamma,\bar{q}^\gamma)}
\end{align*}

\noindent where $\gamma$ runs over the set of fixed points; $f$ is the mean-field free energy that can be derived from eq. (\ref{eq:mf:FE}). These response terms allows one also to compute the skew-symmetric rotation generators of the visible and hidden singular vectors of $\bm{w}$ through
\begin{align*}
  \Omega_{\alpha\beta}^u &= \frac{w_\beta}{w_\alpha^2-w_\beta^2}\bigl(\langle s_\alpha \tau_\beta \rangle_{\rm data}-\langle s_\alpha \tau_\beta \rangle_{\rm {\mathcal H}}\bigr)
  +\frac{w_\alpha}{w_\alpha^2-w_\beta^2}\bigl(\langle s_\beta \tau_\alpha \rangle_{\rm data}-\langle s_\beta \tau_\alpha \rangle_{\rm {\mathcal H}}\bigr),\\[0.2cm]
  \Omega_{\alpha\beta}^v &= \frac{w_\alpha}{w_\alpha^2-w_\beta^2}\bigl(\langle s_\alpha \tau_\beta \rangle_{\rm data}-\langle s_\alpha \tau_\beta \rangle_{\rm {\mathcal H}}\bigr)
  +\frac{w_\beta}{w_\alpha^2-w_\beta^2}\bigl(\langle s_\beta \tau_\alpha \rangle_{\rm data}-\langle s_\beta \tau_\alpha \rangle_{\rm {\mathcal H}}\bigr).
\end{align*}

\begin{figure}
    \centering
    \includegraphics[scale=1.1]{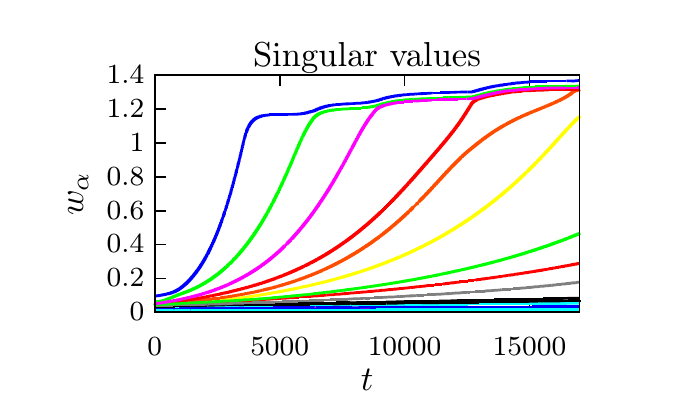}
    \includegraphics[scale=1.1]{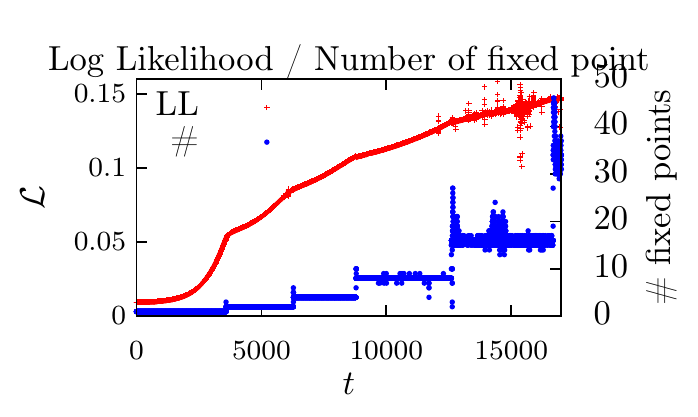}
    \includegraphics[scale=1.2]{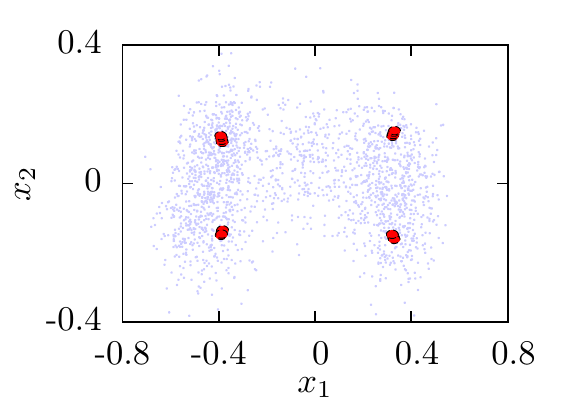}
    \includegraphics[scale=1.2]{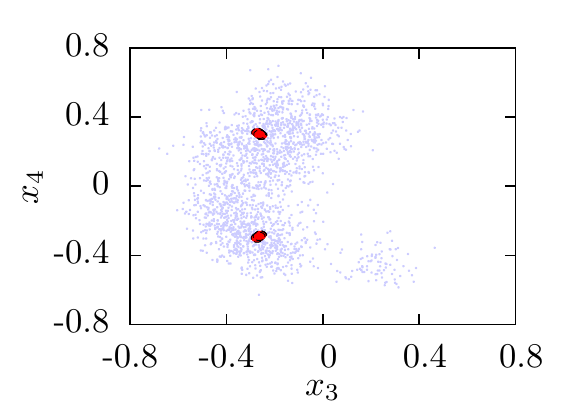}
    \caption{\textbf{Top panel:} Results for a RBM of size $(N_v,N_h)=(1000,500)$ learned on a synthetic dataset of $10^4$ samples having $20$ clusters randomly located on a sub-manifold of dimension $d=15$. The learning curve for the eigemodes $w_\alpha$ (left) and the associated likelihood function (right-red) together with the number of obtained fixed point at each epoch. We can see that, before the first eigenvalue is learned there is one single fixed point, then as modes are learned, the number of fixed points increases. \textbf{Bottom panel:} Results for a RBM of size $(N_v,N_h)=(100,50)$ learned on a synthetic dataset of $10^4$ samples having $11$ clusters randomly defined a sub-manifold of dimension $d=5$. On the left, the scatter plot of the training data together with the position of the fixed points projected on the first two directions of the SVD of $\bm{w}$. On the right, the projection along the third and fourth axis. The results are shown after learning $5$ modes and where $16$ fixed points are found (in fact more than the number of hidden clusters.}
    \label{fig:dynmf}
\end{figure}

With this at hand it is therefore possible to integrate numerically the learning process of the RBM random ensemble defined by (\ref{eq:form_w}) hence given the typical learning trajectory.
If doable in principle with any arbitrary data, this was actually tested in\ucite{decelle2018thermodynamics} on a simple synthetic dataset made of separated clusters. The result is shown on Figure~\ref{fig:dynmf}.

\noindent We see again the different eigenvalues emerging one by one, and that each newly learned eigenvalue is triggering a jump of the likelihood together with a jump in the number of fixed points. At the end of the learning, the obtained mean-field fixed points are located at the center of each cluster of the dataset, as can be seen on the scatter plots. In\ucite{decelle2018thermodynamics}, it is also shown that the behavior is qualitatively similar to what is routinely obtained when performing a standard learning based on PCD.

\subsection{Other mean-field approach}

Other approaches using for instance, message-passing technique such as BP have been developed in order to infer the magnetization of RBM instances. These approaches usually are correct in the limit of weak couplings, and can be used on single instance by iteratively updating a set of messages, here $\mathcal{O}(N_h N_v)$ until convergence (see for instance\ucite{mezard2017mean,huang2015advanced}). In these works, it is shown how BP can be used to infer the magnetization or the free energy in few well-chosen cases. However, as far as practical learning tasks are concerned, it is not clear that this can be used in general when dealing with the ferromagnetic phase, as can be expected when considering structured data. In fact, it has been showed in many works that BP can have very bad convergence properties in a ferromagnetic phase when the underlying factor-graph is not a tree\ucite{lage2013replica} (particularly if the couplings are strong). This would be most probably the case with RBMs. It is also worth mentioning that in the case of the inverse Ising model, BP approaches never manage to succeed because of the convergence problems\ucite{lage2013replica} and the TAP solution was preferred when inferring the couplings\ucite{ricci2012bethe,nguyen2012bethe}. However, some attempts\ucite{huang2016unsupervised,huang2017statistical} using BP and the replica theory on a RBM with one hidden unit were done. In that setting, it is possible to compute the marginal over the weight matrix using BP and therefore to compute its maximum likelihood given some observed datapoints. The results tend to show that, as the number of data increases, the learned features become more localized as is observed in many experiments. Managing to extend this result to the case of many hidden nodes would open the possibility to study the pattern formation using message-passing techniques. An even more recent study\ucite{huang2020variational}, using a variational approach to approximate the posterior distribution of the patterns given the RBM and a dataset shows on artificial data that the patterns are recovered during the learning. Once again, the missing convincing piece in that case is the applicability to real dataset and the ability to samples complex distribution.

\section{Conclusion}

With this review, we strive at showing that not only is RBM part of a hectic field of study, but it is also an intriguing puzzle with pieces which are missing in order to be able to understand the way these models can/could assimilate complex information/more complex information. While the black box nature of the learning process starts to fade away very slowly, there are still many key aspects that we do not understand or master for such simple models. We try to list interesting leads for the future.

\begin{itemize}
    \item \textbf{Learning quality:} Despite the fact that we are maximizing a likelihood function (which can not be computed) it is very hard to obtain a good indicator for comparing two learned RBM. Even if many methods exist to compute the likelihood approximately\ucite{salakhutdinov2008quantitative,krause2020algorithms} the obtained scores are in general not commented in regard to robust statistical analysis. If for very hard cases of image generation, it is easy to compare the results by eye inspection, there are no general method that manage to assess the quality of the samples in terms of how well the learned distribution reproduces the dataset distribution. Some recent work\ucite{yale2020generation} introduced the notion of ``ressemblance'' and ``privacy'' that test the geometric repartition of the true data against the generated samples. This could be a first step defining scores according to different criteria (actually, this problem is not specific to the RBM but concerns actually most of the unsupervised learning models (GANs, VAEs, ...).
    \item \textbf{The number of hidden nodes:} It is striking that we are still unable to have a principled  manner of deciding how many hidden nodes are necessary to learn datasets which are not too complex. For instance, on MNIST, it is possible to learn a machine with only $50$ hidden nodes and it somehow manages to produce decent samples. The understanding on how much hidden nodes are necessary to reach a given sample quality is completely missing. In addition, the number of hidden nodes influences a lot the learning behavior of the machine, again in a way that is not fully understood.
    \item \textbf{The landscape of free energy:} When using statistical mechanics to understand RBMs, the natural question that comes in mind is about the landscape of free energy of the learned machine. It is easy to observe the mean-field fixed point obtained in the ferromagnetic phase and that they do correspond to prototypes of the dataset. Still, we do not know how these many fixed points are organized: are there low free energy paths relating them one from each others? do these paths define a network structure or instead separated clusters of low free energy? 
    \item \textbf{The landscape of learned RBMs:} This is a generic question in Machine Learning : what is the landscape of ``good'' learned machines in parameter space (here the weight matrix). For supervised tasks, some consensus seems to describe a space which is globally flat where all the good model are next to one another. However this is true for deep models, in the case of RBM, apart from the permutation symmetry of the hidden nodes, we have no clue about what this landscape looks like.
    \item \textbf{Link between the dataset and the learned features:} We have seen that in the Gaussian-Gaussian case there is a direct link between the eigen-decomposition of the dataset and the learned features. However, for the non-linear model, we do not understand how the modes of the weight matrix are linked to the dataset, nor to the associated rotation matrices.
\end{itemize}

\bibliographystyle{unsrt}
\bibliography{references}
\end{document}